\documentclass[11pt]{article}
\usepackage{graphicx}
\usepackage{rotating}
\usepackage[utf8]{inputenc}
\usepackage{cmap}
\usepackage{epstopdf}

\begin{document}

\vskip 0.5cm
\centerline{\bf\Large Orbits of 47 Dwarf Satellite Galaxies of the Milky Way in Three}
\centerline{\bf\Large  Models of the Gravitational Potential with Different Masses}
 \bigskip
 \bigskip
  \centerline
 {
  A.~T.~ Bajkova,  V.~V. ~Bobylev
 }
 \bigskip
\centerline{\small \it
Pulkovo Astronomical Observatory, St.-Petersburg, Russia, E-mail: bajkova@gaoran.ru}

 \bigskip
 \bigskip

\begin{abstract}
The analysis of the orbits of 47 dwarf satellite galaxies of the Milky Way, built using three models of the Galactic gravitational potential with different masses, is presented. The models of the Galactic potential were chosen based on the analysis of a large number of the Galaxy mass estimates known from the literature. The astrometric data needed for calculation of the initial positions and velocities (6D phase space) of the dwarf galaxies are also taken from the literature, where the average proper motions were determined on the basis of the data from the Gaia DR2 Catalog. We present the orbits and their properties of all 47 dwarf galaxies obtained by integration for 13.5 Gyr  backward in all three  models of the Galactic potential and give a comparison of the orbital parameters. For each model of the Galactic potential we have identified dwarf galaxies that are not connected gravitationally with the Milky Way.
\end{abstract}


 \section{INTRODUCTION}

Our Galaxy is surrounded by a cloud of dwarf satellite galaxies (McConnachie 2012). More than 60 such satellites are currently known in the region with a radius of about 300 kpc from the Sun (Drlica-Wagner et al. 2019). According to the analysis of the distribution of matter density in the halo and the luminosity function, the expected number of satellite galaxies should be approximately twice as large (Newton et
al. 2018; Nadler et al. 2019). It is likely that the application is more sensitive technology will soon increase the number of faint dwarf satellite galaxies.

The interest in dwarf satellite galaxies of the Milky Way is associated both with their spatial distribution and with their movement. According to modern data on the satellite galaxy population, Deason et al. (2020)  estimated the radius of the Milky Way halo as $292\pm 61$~kpc. To determine such a ``margin'' of the Galaxy, these authors looked for a point of density or line-of-sight velocity fall both in the model distribution of dark matter or stars and in the observed distribution of dwarf satellite galaxies of the Milky Way. In the works of Pawlowski \& Kroupa (2013a,b), attempts were made to detect coherent structures in the distribution and motion of dwarf galaxies of the Local Group. Of interest is the study of the evolution of the entire cloud of Milky Way satellite galaxies, the presence of plumes in the Galaxy and the search for surviving remains from the destruction of dwarf galaxies  (Helmi 2020), the analysis of the evolutionary relationship with globular clusters of the Galaxy (Massari et
al. 2020).

The velocities of satellite galaxies and globular clusters, albeit indirectly (through the dispersion of velocities in the Jeans equation), were used to construct the rotation curve of the Galaxy over a very wide range of distances from the galactic center 0-200 kpc (Bhattacharjee et al. 2014). Such a rotation curve ultimately served as the basis for refining the model of the gravitational potential of the Galaxy, estimating the mass of the Galaxy (Bajkova \& Bobylev 2016;
Sofue 2017; Bobylev et al. 2017), refining the nature of the distribution of invisible matter (Pato \& Iocco 2015).

A reliable study of the kinematics of satellite galaxies requires knowledge of a number of parameters with high accuracy. The distances from variable stars with errors from 5\% to 10\% and line-of-sight velocities, whose errors are independent of distance, are relatively well estimated. The ``weak link'' here is the proper motions, since when moving from the measured velocity components in angular units (mas yr$^{-1}$) to linear (km s$^{-1}$), multiplication by distance occurs. Therefore, the angular movements of dwarf galaxies-satellites of the Milky Way need to know with minimal errors.

For a number of dwarf galaxies, the absolute proper motions were measured using the Hubble Space Telescope (HST) (Pryor et al. 2010, 2015; Kallivayalil et al. 2013). Random errors in determining these proper motions depend on various factors, in particular, from the difference of eras. For example, according to observations on the HST with an epoch difference of 7 years, the proper motions of the Magellanic clouds (LMC and SMC) were measured relative to quasars with errors of about 0.05~mas yr$^{-1}$
(Kallivayalil et
al. 2013). For 11 galaxies such measurements are collected in the work of Pavlovsky, Krupy (2013). A study of the dynamics of the Local Group using this data was performed, for example, by Pawlowski \& Kroupa (2013b), van der Marel
(2015), Bajkova \& Bobylev (2017b).

The appearance of the Gaia DR2 Catalog (Brown et al.
2018; Lindegren et al. 2018) led to a significant increase in the number of dwarf galaxies with calculated high-precision absolute proper motions  (Massari \& Helmi 2018; Fritz et al.
2018, 2019; Simon 2018; Pace \& Li 2019). In the Gaia DR2 Catalog, the average parameter errors depend on the magnitude. For example, the average errors of proper motions range from 0.05~mas yr$^{-1}$ for bright ($G<15^m$) to 1.2~mas yr$^{-1}$ for faint ($G>20^m$) stars.

The closest to us in purposes is the work of Fritz et
al. (2018), based on the data from the Gaia DR2 Catalog and  devoted to study of the orbits of 39 dwarf galaxies, calculated in two potentials with halo of low ($0.8\times10^{12}M_\odot$) and high mass ($1.6\times10^{12}M_\odot$).
Our work is aimed at creating, according to literature, a more complete list of satellite galaxies with data necessary for calculating spatial velocities, and at analysis of their three-dimensional dynamics using models of the Galactic potential, updated recently using modern data on the rotation curve of the Galaxy.
Our list includes 47 dwarfs. The choice of the Galactic gravitational potential models for construction of the orbits is based on the estimates of the mass of the Galaxy obtained by various authors using various objects and methods of analysis. As a result, it was shown that all known estimates lie in the mass range determined by three potential models with masses $M_{G_{(R\leq200~kpc)}}=0.75\times10^{12} M_\odot,1.45\times10^{12} M_\odot$, and $1.90\times10^{12} M_\odot$. We integrate orbits of the dwarfs in all these three models of the Galactic potential and give comparison of the orbit properties.

The work is structured as follows. Section 2 is devoted to the analysis of the Galactic mass estimates from the literature and choosing the Galactic potential models for further dynamic analysis of the dwarf galaxies. Section 3 describes the potential models and equations of motions for orbit integration. Section 4 describes data. In Section 5 we give the results obtained and their analysis. In Conclusions we summarize main results.

\section{Estimates of the Mass of the Galaxy from the Literature}

\begin{figure*}
\begin{center}
\includegraphics[width=0.8\textwidth]{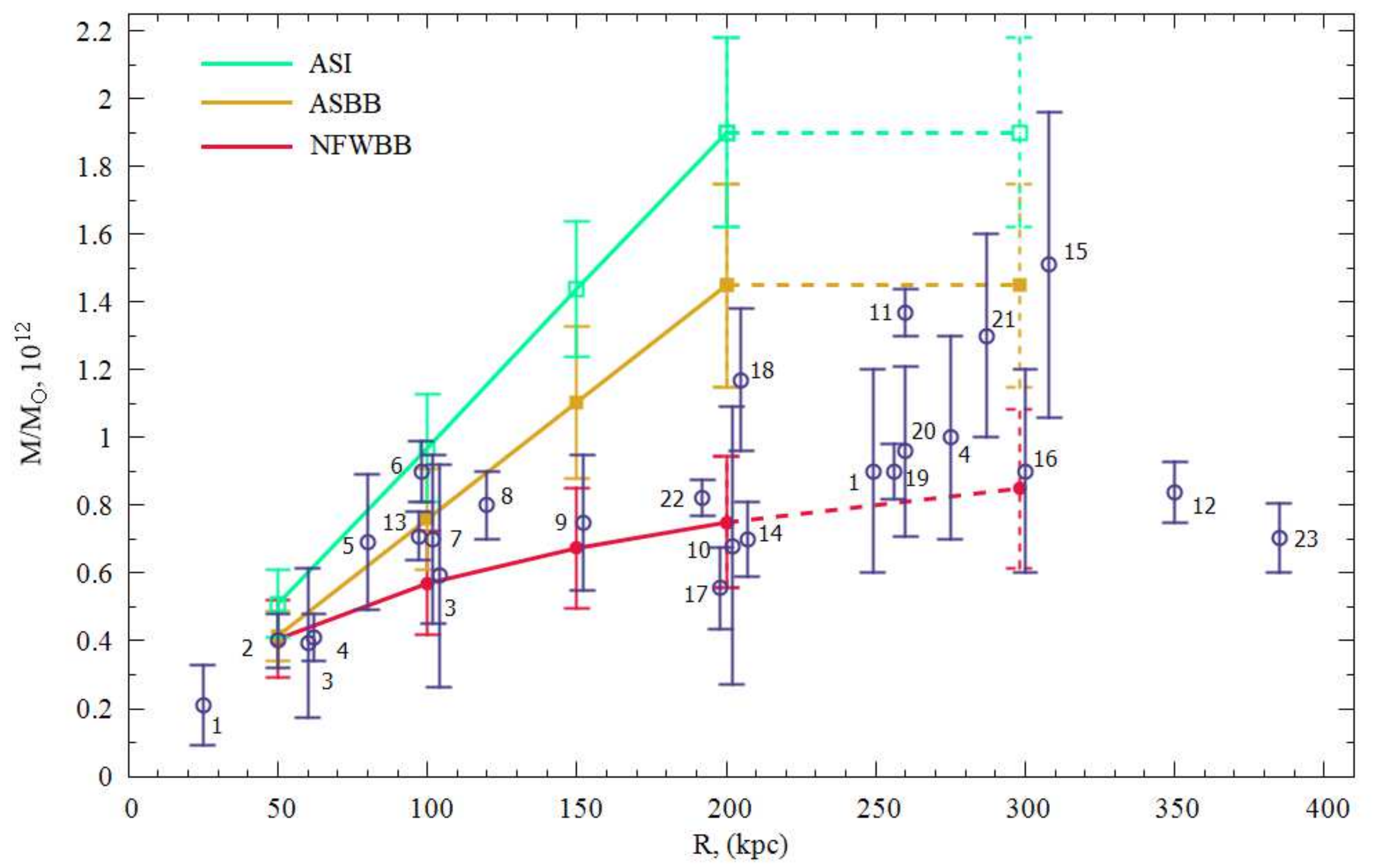}
\caption{\small Estimates of the mass of the Galaxy obtained by various authors:
  1~-- Kafle et al. (2012),        %
  2~-- Deason et al. (2012a),     %
  3~-- Bhattacharjee et al.~(2013), %
  4~-- Xue et al. (2008),         %
  5~-- Gnedin et al. (2010),      %
  6~-- McMillan (2011),         %
  7~-- Dehnen, Binney (1998),      %
  8~-- Battaglia et al. (2005),    %
  9~-- Deason et al. (2012b),     %
 10~-- Bhattacharjee et al.~(2014), %
 11~-- Eadie et al. (2015),         %
 12~-- Karachentsev et al. (2009),  %
 13~-- Fardal et al. (2019),      
 14~-- Eadie et al. (2019),         
 15~-- Fritz et al. (2020),        
 16~-- Watkins et al. (2010),     
 17~-- Gibbons et al. (2014),     
 18~-- Callingham et al. (2019),   
 19~-- Huang et al. (2016),       
 20~-- Patel et al. (2018),      
 21~-- Posti, Helmi (2019),     
 22~-- Ablimit et al. (2020),     
 23~-- Sofue (2012).              
 A description of three potential models (NFWBB, ASBB, ASI) is given in the text.
}
\label{fmass}
\end{center}
\end{figure*}

So, to select the most probable models of the Galactic potential in order to integrate the orbits of dwarf satellite galaxies, we collected the Galaxy mass estimates from various literary sources obtained by independent methods, and plotted them on one graph (Fig.~\ref{fmass}).

Most of the presented mass estimates are based on the construction of the rotation curve of the Galaxy from one or another data. To construct the rotation curve of the Galaxy, giant stars that are visible at large distances are often used. Other important objects for solving this problem are Cepheids, stars of the RR Lyra type, globular clusters, or dwarf satellite galaxies of the Milky Way.

In a paper by Dehnen \& Binney (1998), a Galaxy rotation curve was constructed up to a distance of $R=100$~kpc using a variety of data. In particular, data on the velocities of the H II, Cepheids, and dwarf galaxies were used. The data on proper motions were taken from the HIPPARCOS Catalog (1997). A three-component axisymmetric model of the gravitational potential was used to determine the law of density distribution in the Galaxy.
The same method was later applied by McMillan (2011).
In the work of Kafle et al. (2012), the line-of-sight velocities of more than 4,500 blue branch giants with distances $R$ up to $60$~kpc were taken. These are stars from the SDSS Catalog (Aihara et al. 2011) with line-of-sight velocities from the SEGUE Catalog (Yanny et al. 2009). To fit the rotation curve, they used a three-component axisymmetric model of the Galactic potential, including  the bulge, disk, and halo. The mass of the Galaxy was estimated on the basis of the Jeans equation by the radial velocity dispersion profile, the anisotropy profile, and the density distribution law. Fig.~\ref{fmass} gives two estimates by these authors --- for $R = 25$~kpc and $R=249$~kpc.
According to the data on blue giants of the horizontal branch from the SDSS Catalog, Deason et al. (2012a) obtained an estimate $M_{G_{(R\leq50~kpc)}}=(0.42\pm0.04)\times10^{12}M_\odot$. In another work by these authors (Deason et al., 2012b), $M_{G_{(R\leq150~kpc)}}=(0.5-1.0)\times10^{12}M_\odot$ was found from farther blue giants of the horizontal branch.
Xue et al. (2008) analyzed the line-of-sight velocities of the blue giants, which were located at distances $R<60$~kpc. A three-component model of the Galactic potential was constructed in which the halo of dark matter is presented in the Navarro-Frank-White form  (Navarro et al. 1997). These authors estimated the virial mass of the Galaxy as $M_{G_{(R\leq R_{vir})}}=(1.0^{+0.3}_{-0.2})\times10^{12}M_\odot$, where the value of the virial radius was $R_{vir}=275^{+23}_{-20}$~kpc.
Two estimates of these authors are given in Fig.~\ref{fmass} --- for $R=55$~kpc and $R=275$~kpc.
Huang et al. (2016) used about 16,000 giants from redshifts LEGUE Catalog  (Deng et al. 2012) and about 5700 giants of the spectral class K from the SDSS Catalog. These authors estimated the virial mass of the Galaxy as $M_{G_{(R\leq R_{vir})}}=(0.90^{+0.07}_{-0.08})\times10^{12}M_\odot$, where the value of the virial radius is $R_{vir}=255.7^{+7.7}_{-7.7}$~kpc.
Fardal
et al. (2019) used a sample of stars of the RR Lyra type from the Pan-STARRS1 Catalog (Hernitschek et al. 2017) to trace the Sagittarius Stream (Sgr Stream).
As a result, these authors found
  $M_{G_{(R\leq60~kpc)}}=(0.41\pm0.04)\times10^{12}M_\odot$ and
  $M_{G_{(R\leq100~kpc)}}=(0.71\pm0.07)\times10^{12}M_\odot$.
Gibbons et al. (2014) from the analysis of Sagittarius Stream, received an estimate
$M_{G_{(R\leq100~kpc)}}=(0.41\pm0.04)\times10^{12}M_\odot$, which was extrapolated to 200 kpc, and found $M_{G_{(R\leq200~kpc)}}=(0.56\pm0.12)\times10^{12}M_\odot$.
Gnedin et al. (2010) considered a galactic rotation curve up to distances $R=80$~kpc constructed using the line-of-sight velocities of distant halo stars.
Battaglia et al. (2005) used line-of-sight velocities of 240 halo stars.
In the works of Bhattacharjee et al.
(2013), line-of-sight velocities of globular clusters and dwarf galaxies were used to construct a galactic rotation curve at large distances. Bhattacharjee et al. (2014) found an estimate $M_{G_{(R\leq200~kpc)}}=(0.68\pm0.41)\times10^{12}M_\odot$.
Eadie et al.
(2015) estimated the mass of the Galaxy
as $M_{G_{(R\leq260~kpc)}}=(1.37\pm0.07)\times10^{12}M_\odot$
according to data on globular clusters and dwarf galaxies located up to $R=260$ kpc.
Eadie \& Juri\'c (2019) used the latest data on globular clusters.
Fritz et al. (2020) obtained a virial estimate of the mass of the Galaxy $M_{G_{(R\leq R_{vir})}}=(1.51^{+0.45}_{-0.40})\times10^{12}M_\odot$ from data on dwarf galaxies, where $R_{vir}=308$~kpc. Their work provides a fairly complete overview of the results of determining the mass of the Galaxy by various authors.
Watkins et al. (2010) used data on halo stars, globular clusters and dwarf galaxies and found $M_{G_{(R\leq300~kpc)}}=(0.9\pm0.3)\times10^{12} M_\odot$.
Callingham et al. (2019) obtained $M_{G_{(R\leq200~kpc)}}=(1.17^{+0.21}_{-0.15})\times10^{12} M_\odot$
from data on dwarf galaxies satellites of the Milky Way.
Patel et al. (2018) found $M_{G_{(R\leq260~kpc)}}=(0.96^{+0.29}_{-0.28})\times10^{12} M_\odot$ from data on dwarf galaxies satellites of the Milky Way.
Posti \& Helmi (2019) obtained a virial estimate of the mass of the Galaxy
as  $M_{G_{(R\leq R_{vir})}}=(1.3\pm0.3)\times10^{12} M_\odot$, where
   $R_{vir}=287^{+22}_{-25}$~kpc.
 Ablimit et al. (2020) used about 3,500 classical Cepheids with data from the OGLE catalogs (Optical Gravitational Lensing Experiment)  (Udalski
et al. 1997), ASAS (All Sky Automated Survey)  (Pojmanski 2002) and Gaia DR2  (Prusti et al. 2016; Brown et al.
2018) to construct the rotation curve of the Galaxy.
These authors obtained a virial estimate of the mass of the Galaxy
  $M_{G_{(R\leq192~kpc)}}=(0.822\pm0.052)\times10^{12} M_\odot$.
Based on a three-component model of the Galactic potential in which the halo of dark matter is presented in the Navarro-Frank-White form  (Navarro et al. 1997),  Sofue (2012) obtained a virial estimate of the mass of the Galaxy
$M_{G_{(R\leq 385~kpc)}}=(0.70\pm0.10)\times10^{12}M_\odot$, where half of the distance to the Andromeda nebula was chosen as an estimate of the virial radius.
Karachentsev et al. (2009) estimated the total mass of the Local Group $(19\pm0.2)\times10^{12}M_\odot$ and the ratio of the Milky Way and M31 masses as 4:5. For this assessment, the braking effect of the local Hubble stream and data on the distances and line-of-sight velocities of galaxies in the vicinity of the Local Group were used. The total mass of the Galaxy obtained by this independent method is  $M_{G_{(R\leq350~kpc)}}=(0.84\pm0.09)\times10^{12}M_\odot$.
Note that a similar independent analysis of Galactic mass estimates is given in a recent paper by Wang et al. (2019).

Fig.~\ref{fmass} shows three functions of the mass of the Galaxy versus the distance $R$, corresponding to three chosen by us models of the Galactic potential, which cover, taking into account the error bars, almost the entire range of given estimates of the Galaxy mass. The lower curve corresponds to the model of the potential with a dark halo in the form of Navarro-Frenk-White (Navarro et al. 1997), modified by Bajkova \& Bobylev (2016).  The mass of the Galaxy according to this model is $M_{G_{(R \leq 200 kpc)}}=0.75\pm0.19\times10^{12}M_\odot$. For brevity, we designate this model as NFWBB. The middle curve corresponds to the Allen \& Santill\'an (1991) model of the potential, also modified in Bajkova \& Bobylev (2016). The mass of the Galaxy in this model, designated as ASBB, is $M_{G_{(R \leq 200 kpc)}}=1.45\pm0.30\times10^{12}M_\odot$. And finally, the third model, with the largest mass of the Galaxy, equal to
$M_{G_{(R\leq 200~kpc)}}=1.9^{+2.4}_{-0.8}\times10^{12}M_\odot$, corresponds to the upper curve. The third model is the Allen \& Santill\'an (1991) model, modified by Irrgang et al. (2013). We designate it as ASI model. More detail description of these models is given in Section~3.1.

 \section{METHOD}
 \subsection{The Models of the Galactic Gravitational Potential}
In all of the models here, the axially symmetric gravitational potential of the Galaxy is
represented as the sum of three components --- the central,
spherical bulge $\Phi_b(r(R,Z))$, the disk $\Phi_d(r(R,Z))$, and
the massive, spherical dark-matter halo $\Phi_h(r(R,Z))$:
 \begin{equation}
 \begin{array}{lll}
  \Phi(R,Z)=\Phi_b(r(R,Z))+\Phi_d(r(R,Z))+\Phi_h(r(R,Z)).
 \label{pot}
 \end{array}
 \end{equation}
Here, we use a cylindrical coordinate system ($R,\psi,Z$) with
its origin at the Galactic center. In Cartesian coordinates
$(X,Y,Z)$ with their origin at the Galactic center, the distance
to a star (the spherical radius) is $r^2=X^2+Y^2+Z^2=R^2+Z^2$.
The gravitational potential is expressed in units of 100 km$^2$ s$^{-2}$, distances in kpc,
masses in units of the mass of the Galaxy, $M_{gal}=2.325\times
10^7 M_\odot$, and the gravitational constant is taken to be
$G=1.$

We express the potentials of the bulge, $\Phi_b(r(R,Z))$, and
disk, $\Phi_d(r(R,Z))$, in the form suggested by Miyamoto \&
Nagai (1975):
 \begin{equation}
  \Phi_b(r)=-\frac{M_b}{(r^2+b_b^2)^{1/2}},
  \label{bulge}
 \end{equation}
 \begin{equation}
 \Phi_d(R,Z)=-\frac{M_d}{\Biggl[R^2+\Bigl(a_d+\sqrt{Z^2+b_d^2}\Bigr)^2\Biggr]^{1/2}},
 \label{disk}
\end{equation}
where $M_b, M_d$ are the masses of these components, and $b_b,
a_d, b_d$ are the scale parameters of the components in kpc.

 For description of the halo component, we used (i) the expression in Navarro-Frenk-White (NFW) form presented in Navarro et al. (1997):
 \begin{equation}
  \Phi_h(r)=-\frac{M_h}{r} \ln {\Biggl(1+\frac{r}{a_h}\Biggr)},
 \label{halo-III}
 \end{equation}
where $M_h$ is the mass, $a_h$ is the scale length,
and (ii) the expression for Allen \& Santill\'an (1991) model (AS) derived by Irrgang et al. (2013) from the expression for the halo mass  (Allen \& Martos 1986):
 \begin{equation}
 \renewcommand{\arraystretch}{2.8}
  \Phi_h(r) = \left\{
  \begin{array}{ll}
  \displaystyle
  \frac{M_h}{a_h}\biggl( \frac{1}{(\gamma-1)}\ln
  \biggl(\frac{1+(r/a_h)^{\gamma-1}}{1+(\Lambda/a_h)^{\gamma-1}}\biggr)-\\
    \displaystyle
    \qquad\qquad-\frac{(\Lambda/a_h)^{\gamma-1}}{1+(\Lambda/a_h)^{\gamma-1}}\biggr),
  \quad\textrm{if }   r\leq \Lambda & \\
  \displaystyle
  -\frac{M_h}{r} \frac{(\Lambda/a_h)^{\gamma}}{1+(\Lambda/a_h)^{\gamma-1}},
  \qquad\qquad\textrm{if }  r>\Lambda, &
  \end{array}\right.
 \label{halo-I}
 \end{equation}
where $M_h$ is the mass, $a_h$ is the scale length, the Galactocentric distance is $\Lambda=200$ kpc, and the dimensionless coefficient $\gamma=2.0$.

The first model (NFWBB) of the Galactic potential, considered in this work, is the NFW model modified
in work of Bajkova \& Bobylev (2016) by fitting of the model parameters to data on HI,
maser sources and Galactic objects from Bhattacharjee et al. (2014) at distances $R$ within $\sim200$
kpc. In addition the constraints (Irrgang et al. 2013) on
the local dynamical matter density $\rho_\odot=0.1M_\odot$~pc$^{-3}$ and the force acting perpendicularly to the Galactic plane $|K_{z=1.1}|/2\pi G=77M_\odot$~pc$^{-2}$ were used.

The second model (ASBB) is the AS model modified also in work of Bajkova \& Bobylev (2016) using the same data and the same local constraints.

Note that among six  models of the Galactic potential summarized in Bajkova \& Bobylev (2017b) the NFWBB model ensures the best fit to the data.

The third model (ASI), considered in this work, is the AS model modified by Irrgang et al. (2013) using data on HI, CO and maser sources at distances up to 20 kpc and the same local constraints as in previous cases.

\begin{figure*}
\begin{center}
   \includegraphics[width=0.9\textwidth,angle=0]{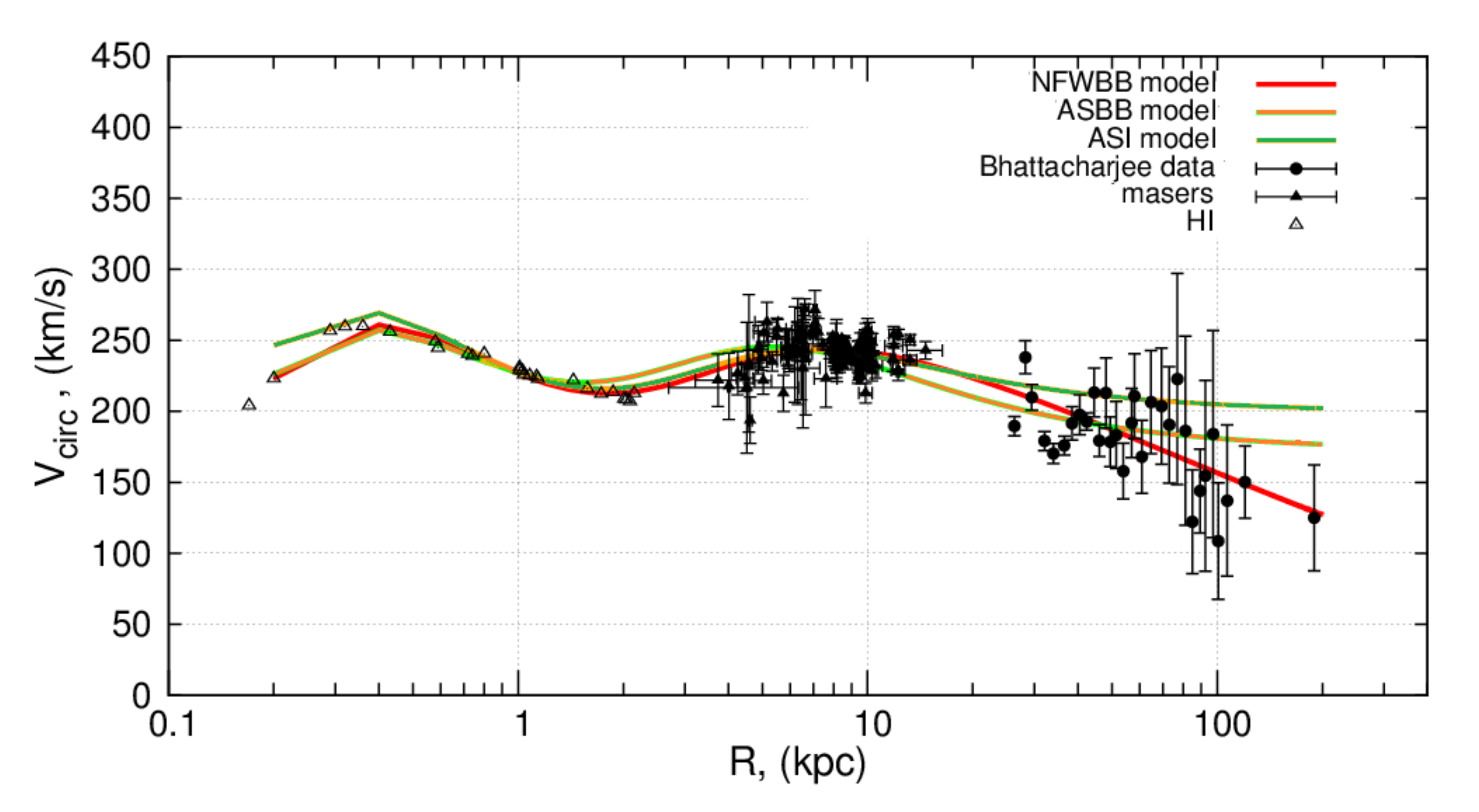}
  \caption{Rotation curves of three potential models (NFWBB, ASBB, ASI)}
\label{rotc}
\end{center}
\end{figure*}

Parameters of these three models are given in Table \ref{t:prod}, where local parameters $R_\odot$, $V_\odot$ and the total mass of the Galaxy within a sphere of radius 200 kpc are indicated as well. The mass distribution is shown in Fig.\ref{fmass}. Corresponding rotation curves up to $R=200$ kpc are shown in Fig.\ref{rotc}.

 \begin{table}
 \begin{center}
 \caption[]
 {\small\baselineskip=1.0ex
 Parameters of the models of the Galactic potential\\
  ($M_0=2.325\times10^7 M_\odot$)}
  \bigskip
 \label{t:prod}
 \begin{tabular}{|c|r|r|r|}\hline
 Parameter     &  NFWBB           & ASBB   & ASI                                  \\\hline
 $M_b$ [$M_0$] &    443$\pm$27    & 386$\pm$10       & 409$\pm$63                 \\
  $M_d$ [$M_0$] &   2798$\pm$84    & 3092$\pm$62      & 2856$^{+376}_{-202}$      \\
  $M_h$ [$M_0$] &  12474$\pm$3289  & 452$\pm$83       & 1018$^{+27933}_{-603}$    \\
  $b_b$ [kpc]   & 0.267$\pm$0.009 & 0.249$\pm$0.006& 0.23$\pm$0.03            \\
  $a_d$ [kpc]   &   4.40$\pm$0.73  & 3.67$\pm$0.16    & 4.22$^{+0.53}_{-0.99}$    \\
  $b_d$ [kpc]   & 0.308$\pm$0.005& 0.305$\pm$0.003& 0.292$^{+0.020}_{-0.025}$ \\
  $a_h$ [kpc]   &    7.7$\pm$2.1   & 1.52$\pm$0.18    & 2.562$^{+25.963}_{-1.419}$\\
  $\Lambda$ [kpc]&  --- & 200      & 200$^{+0}_{-83}$                             \\
 $\gamma$       &  --- &  2.0     & 2.0                                           \\\hline
 $R_\odot$ [kpc]    &  8.30  &  8.30  & 8.33\\\hline
 $V_\odot$ [km s$^{-1}$] & 243.9 & 239.0& 242.0 \\\hline
 $M_{G_{(R \leq 200 kpc)}}$ & 0.75$\pm$0.19 & 1.45$\pm$0.30 & 1.90$^{+2.4}_{-0.8}$ \\
  $[10^{12} M_\odot]$     &               &               &                      \\\hline
  \end{tabular}
  \end{center}
  \end{table}

 \subsection{Integrating of the Orbits and Computing of their Parameters}
The equation of motion of a test particle in an axially symmetric
gravitational potential can be obtained from the Lagrangian of the
system $\pounds$ (see Appendix A in  Irrgang et al.
(2013)):
\begin{equation}
 \begin{array}{lll}
 \pounds(R,Z,\dot{R},\dot{\psi},\dot{Z})=\\
 \qquad0.5(\dot{R}^2+(R\dot{\psi})^2+\dot{Z}^2)-\Phi(R,Z).
 \label{Lagr}
 \end{array}
\end{equation}
Introducing the canonical moments
\begin{equation}
 \begin{array}{lll}
    p_{R}=\partial\pounds/\partial\dot{R}=\dot{R},\\
 p_{\psi}=\partial\pounds/\partial\dot{\phi}=R^2\dot{\psi},\\
    p_{Z}=\partial\pounds/\partial\dot{Z}=\dot{Z},
 \label{moments}
 \end{array}
\end{equation}
we obtain the Lagrangian equations in the form of a system of six
first-order differential equations:
 \begin{equation}
 \begin{array}{llllll}
 \dot{R}=p_R,\\
 \dot{\psi}=p_{\psi}/R^2,\\
 \dot{Z}=p_Z,\\
 \dot{p_R}=-\partial\Phi(R,Z)/\partial R +p_{\psi}^2/R^3,\\
 \dot{p_{\psi}}=0,\\
 \dot{p_Z}=-\partial\Phi(R,Z)/\partial Z.
 \label{eq-motion}
 \end{array}
\end{equation}
We integrated Eqs. (\ref{eq-motion}) using a fourth-order Runge-Kutta algorithm.

The Sun's peculiar velocity with respect to the Local Standard of
Rest was taken to be
$(u_\odot,v_\odot,w_\odot)=(11.1,12.2,7.3)\pm(0.7,0.5,0.4)$ km s$^{-1}$
(Sch\"onrich et al.
2010). Here, we use the heliocentric velocities in a moving
Cartesian coordinate system with $u$ directed towards the Galactic
center, $v$ in the direction of the Galactic rotation, and $w$
perpendicular to the Galactic plane and directed towards the north
Galactic pole.

Let the initial positions and space velocities (6D phase space) of a test particle
in the heliocentric coordinate system be
$(x_o,y_o,z_o,u_o,v_o,w_o)$. The initial positions ($X,Y,Z$) and velocities ($U,V,W$)
of the test particle in Galactic Cartesian coordinates are then
given by the formulas:
\begin{equation}
 \begin{array}{llllll}
 X=R_\odot-x_o, Y=y_o, Z=z_o+h_\odot,\\
 R=\sqrt{X^2+Y^2},\\
 U=u_o+u_\odot,\\
 V=v_o+v_\odot+V_\odot,\\
 W=w_o+w_\odot,
 \label{init}
 \end{array}
\end{equation}
where $R_\odot$ and $V_\odot$ are the Galactocentric distance and the
linear velocity of the Local Standard of Rest around the Galactic
center, $h_\odot=16$~pc  (Bobylev \& Bajkova 2016) is the height of the Sun above the Galactic plane,$\Pi$ and $\Theta$ are radial and tangential velocities respectively.

Below in Table \ref{t:prop} the following orbital parameters of dwarf satellite galaxies are given:

\noindent (1)initial distance of the dwarf galaxy from the Galactic center $d_{GC}$:
\begin{equation}
d_{GC}=\sqrt{X^2+Y^2+Z^2};
\end{equation}

\noindent (2)radial velocity $\Pi$:
\begin{equation}
 \Pi=-U \frac{X}{R}+V \frac{Y}{R};
 \end{equation}

\noindent (3)tangential velocity $\Theta$:
\begin{equation}
 \Theta=U \frac{Y}{R}+V \frac{X}{R};
 \end{equation}

\noindent (4)total 3D velocity $V_{tot}$:
\begin{equation}
V_{tot}=\sqrt{\Pi^2+\Theta^2+W^2};
 \end{equation}

\noindent (5)apocentric distance (apo) of the orbit;

\noindent (6)pericentric distance (peri) of the orbit;

\noindent (7)the eccentricity (ecc) of the orbit:
\begin{equation}
ecc=\frac{apo-peri}{apo+peri};
 \end{equation}

\noindent (8)the components of the angular momentum:
\begin{equation}
 L_X=Y\times W-Z\times V;
 \end{equation}

 \begin{equation}
 L_Y=Z\times U-X\times W;
 \end{equation}

\begin{equation}
 L_Z=X\times V-Y\times U;
 \end{equation}

\noindent (9)inclination of the orbit $\theta$:
\begin{equation}
\theta=\arccos(\frac{L_Z}{L}),
\end{equation}
where $L=\sqrt{L_X^2+L_Y^2+L_Z^2}$;

\noindent (10)period of the orbit $T_r$;

\noindent (11)total energy $E$:
\begin{equation}
 E= \Phi(R,Z)+\frac{V_{tot}^2}{2}.
 \end{equation}

 The uncertainties of the parameters were calculated by the Monte-Carlo simulation using 100 iterations, taking into account the uncertainties in the initial coordinates and velocities of dwarf galaxies, as well as errors in the peculiar velocity of the Sun.

\section{DATA}
The average values of the proper motions of dwarf galaxies were calculated according to the Gaia DR2 Catalog in the works of various authors (Simon 2018; Massari \& Helmi
2018; Kallivayalil et al. 2018; Pace \& Li 2019; Fritz et al.
2018, 2019).  Fritz et al. (2018) compared their calculations with a number of the indicated works and concluded that in most cases there is good agreement of the results within the limits of errors at the level of 1 sigma. These authors noted a difference of more than 2.5 sigma for the Cra II, and Tuc III galaxies compared with the definitions of Kallivayalil et
al. (2018), possibly related to the influence of background stars. They also noted a very faint Segue 2 galaxy, whose foreground stars interfere with reliable selection of members.

Our main worklist is the work of Fritz et al. (2018), from which data on 39 galaxies were taken. For the galaxy UMi I, there are average values of proper motions in both Fritz et al. (2018) and Pace et al. (2020). These values are consistent with each other within the error of their determination. To derive the average values of the proper motions of the galaxy UMi I Fritz et al. (2018) used 137 stars common with the Gaia DR2 catalog. We preferred data from the work of Pace et al. (2020), where 892 stars (with a probability of belonging to the galaxy of more than 0.95) were refined with an analysis of the line-of-sight velocities of stars and photometric data from several surveys.

For four other galaxies, ColI, HorII, PhxII and RetIII, from Fritz et al. (2019), not only new estimates of proper motions were taken, but also distance estimates, as well as line-of-sight velocities. Finally, the list was supplemented by new estimates of the proper motions of the following galaxies: LMC and SMC (Helmi et al. 2018), Boo III  (Massari \&
Helmi 2018) and SgrII  (Longeard et al. 2020).

The worklist thus formed (see Table \ref{t:data}) contains measurement data and their errors on 47 dwarf galaxies. The initial position and velocity coordinates $(x_o,y_o,z_o,u_o,v_o,w_o)$ calculated from these data and used for integrating of the orbits are given in Table \ref{t:crd}. Uncertainties in the initial coordinates were calculated using Monte-Carlo simulation (1000 iterations) taking into account the measurement errors. The uncertainty in the dwarf galaxies heliocentric distances $d$ was adopted as 7\% from $d$.

\begin{table*}
 \begin{center}
 \caption[]
 {\small\baselineskip=1.0ex
 Astrometric data for Dwarf galaxies
  }
  \bigskip
 \label{t:data}
 \begin{small}
 \begin{tabular}{|l|r|r|r|r|r|r|r|r|r|}\hline
 Name & $\alpha$&$\delta$&$d$&$\mu_\alpha^*$&$\epsilon_{\mu_\alpha^*}$&$\mu_\delta$&$\epsilon_{\mu_\delta}$&
 $V_r$&$\epsilon_{V_r}$         \\
    & [deg]& [deg]&[kpc]&[mas yr$^{-1}$]&[mas yr$^{-1}]$&[mas yr$^{-1}]$&[mas yr$^{-1}]$&[km s$^{-1}]$&[km s$^{-1}]$\\\hline
AquII        &    338.48&     -8.67&    107.90&    -0.252&     0.063&     0.011&     0.063&     -71.1&       2.5\\
BooI         &    210.00&     14.50&     66.38&    -0.554&     0.035&    -1.111&     0.035&      99.0&       2.1\\
BooII        &    209.50&     12.85&     41.17&    -2.686&     0.056&    -0.530&     0.056&    -117.0&       5.2\\
BooIII       &    209.28&     26.78&     50.00&    -1.210&     0.130&    -0.920&     0.000&     197.5&       3.8\\
CVenI        &    202.01&     33.56&    211.27&    -0.159&     0.035&    -0.067&     0.035&      30.9&       0.6\\
CVenII       &    194.29&     34.32&    160.40&    -0.342&     0.056&    -0.473&     0.056&    -128.9&       1.2\\
CarI         &    100.44&    -49.05&    102.93&     0.485&     0.035&     0.131&     0.035&     229.1&       0.1\\
CarII        &    114.11&    -56.00&     36.20&     1.867&     0.035&     0.082&     0.035&     477.2&       1.2\\
CarIII       &    114.63&    -56.10&     27.80&     3.046&     0.057&     1.565&     0.057&     284.6&       3.4\\
CBerI        &    186.75&     23.90&     41.76&     0.471&     0.035&    -1.716&     0.035&      98.1&       0.9\\
ColI         &     82.86&    -27.97&    182.00&     0.250&     0.240&    -0.440&     0.330&     153.7&       5.0\\
CraI         &    174.07&     -9.12&    144.97&    -0.045&     0.063&    -0.165&     0.063&     149.3&       1.2\\
CraII        &    177.31&    -17.59&    117.50&    -0.184&     0.035&    -0.106&     0.035&      87.5&       0.4\\
DraI         &    260.06&     57.92&     79.18&    -0.012&     0.035&    -0.158&     0.035&    -291.0&       0.1\\
DraII        &    238.20&     64.57&     20.00&     1.242&     0.057&     0.845&     0.057&    -347.6&       1.8\\
EriII        &     56.09&    -42.47&    362.90&     0.159&     0.053&     0.372&     0.053&      75.6&       2.4\\
FnxI         &     39.97&    -33.49&    138.77&     0.374&     0.035&    -0.401&     0.035&      55.3&       0.3\\
GruI         &    344.18&    -49.84&    119.75&    -0.261&     0.046&    -0.437&     0.046&    -140.5&       2.4\\
HerI         &    247.76&     12.79&    134.71&    -0.297&     0.035&    -0.329&     0.035&      45.0&       1.1\\
HorI         &     43.88&    -53.88&     79.00&     0.891&     0.058&    -0.550&     0.058&     112.8&       2.6\\
HorII        &     49.13&    -49.98&     76.00&     1.520&     0.250&    -0.470&     0.390&     168.7&      12.9\\
HyaII        &    185.43&    -30.01&    134.00&    -0.416&     0.061&     0.134&     0.061&     303.1&       1.4\\
HyiI         &     37.39&    -78.69&     27.64&     3.733&     0.035&    -1.605&     0.035&      80.4&       0.6\\
LeoI         &    152.12&     12.31&    269.07&    -0.086&     0.035&    -0.128&     0.035&     282.5&       0.5\\
LeoII        &    168.37&     22.15&    224.39&    -0.025&     0.035&    -0.173&     0.035&      78.0&       0.1\\
LeoIV        &    173.24&      0.53&    154.23&    -0.590&     0.059&    -0.449&     0.059&     132.3&       1.4\\
LeoV         &    172.79&      2.22&    173.09&    -0.097&     0.057&    -0.628&     0.057&     173.3&       3.1\\
LMC          &     80.89&    -68.24&     50.00&     1.850&     0.030&     0.234&     0.030&     262.2&       3.4\\
PheII        &    355.00&    -53.59&     80.00&     0.500&     0.120&    -1.160&     0.140&      32.4&       3.8\\
PhxI         &     27.78&    -43.56&    418.87&     0.079&     0.040&    -0.049&     0.040&     -21.2&       1.0\\
PisII        &    344.63&      5.95&    183.01&    -0.108&     0.061&    -0.586&     0.061&    -226.5&       2.7\\
RetII        &     53.93&    -53.95&     30.00&     2.398&     0.035&    -1.319&     0.035&      62.8&       0.5\\
RetIII       &     56.36&    -59.55&     92.00&    -0.390&     0.530&    -0.320&     0.630&     274.2&       7.5\\
SgrI         &    283.76&    -29.52&     26.00&    -2.736&     0.035&    -1.357&     0.035&     140.0&       2.0\\
SgrII        &    298.17&    -21.93&     67.00&    -0.650&     0.100&    -0.880&     0.120&    -177.3&       1.2\\
SclI         &     15.03&    -32.33&     84.77&     0.084&     0.035&    -0.133&     0.035&     111.4&       0.1\\
Seg1         &    151.76&     16.07&     23.02&    -1.697&     0.035&    -3.501&     0.035&     208.5&       0.9\\
Seg2         &     34.82&     20.18&     36.09&     1.656&     0.045&     0.135&     0.045&     -39.2&       2.5\\
SMC          &     13.16&    -71.20&     64.00&     0.797&     0.030&    -1.220&     0.030&     145.6&       0.6\\
SxtI         &    153.26&     -0.39&     85.74&    -0.438&     0.035&     0.055&     0.035&     224.2&       0.1\\
TriII        &     33.32&     36.18&     30.00&     0.588&     0.051&     0.554&     0.051&    -381.7&       1.1\\
TucII        &    342.98&    -57.43&     57.00&     0.910&     0.035&    -1.159&     0.035&    -129.1&       3.5\\
TucIII       &    359.15&    -58.40&     25.00&    -0.025&     0.035&    -1.661&     0.035&    -102.3&       2.0\\
UMaI         &    158.72&     51.92&     97.34&    -0.683&     0.035&    -0.720&     0.035&     -55.3&       1.4\\
UMaII        &    132.87&     63.13&     34.88&     1.691&     0.035&    -1.902&     0.035&    -116.5&       1.9\\
UMiI         &    227.27&     67.24&     76.22&    -0.152&     0.014&     0.064&     0.013&    -244.7&       0.3\\
Wil1         &    162.34&     51.05&     38.28&     0.199&     0.053&    -1.342&     0.053&     -12.3&       2.5\\
\hline
 \end{tabular}
 \end{small}
  \end{center}
  \end{table*}

\begin{table*}
 \begin{center}
 \caption[]
 {\small\baselineskip=1.0ex
 Dwarf galaxies positions and velocities from the Sun, not corrected for the solar motion or the LSR
  }
  \bigskip
 \label{t:crd}
 \begin{small}
 \begin{tabular}{|l|r|r|r|r|r|r|}\hline
 Name & $x_o$ [kpc]&$y_o$ [kpc]&$z_o$ [kpc]&$u_o$ [km/s]&$v_o$ [km/s]&$w_o$ [km/s] \\
\\\hline
AquII       & $ 38.0^{+2.6}_{-2.8}$& $ 53.3^{+3.7}_{-3.9}$& $-85.8^{+6.2}_{-5.9}$& $ 80.7^{+30.3}_{-31.7}$& $ -3.3^{+27.2}_{-29.7}$& $123.2^{+19.3}_{-21.3}$\\
BooI        & $ 23.0^{+1.6}_{-1.8}$& $ -1.4^{+0.1}_{-0.1}$& $ 62.2^{+4.4}_{-4.7}$& $103.2^{+11.3}_{-11.8}$& $-385.1^{+32.0}_{-28.4}$& $ 58.6^{+5.2}_{-5.0}$\\
BooII       & $ 14.7^{+1.0}_{-1.0}$& $ -2.0^{+0.1}_{-0.1}$& $ 38.4^{+2.6}_{-2.7}$& $-377.2^{+26.4}_{-24.3}$& $-396.1^{+29.8}_{-29.1}$& $ -1.9^{+9.1}_{-9.9}$\\
BooIII      & $ 10.5^{+0.7}_{-0.8}$& $  7.0^{+0.5}_{-0.5}$& $ 48.4^{+3.3}_{-3.5}$& $-37.7^{+23.6}_{-24.0}$& $-317.3^{+31.4}_{-30.2}$& $258.3^{+9.7}_{-9.8}$\\
CVenI       & $ 11.1^{+0.8}_{-0.8}$& $ 35.7^{+2.4}_{-2.5}$& $207.9^{+14.2}_{-14.5}$& $-86.9^{+34.7}_{-35.4}$& $-140.2^{+34.7}_{-35.1}$& $ 60.1^{+6.5}_{-6.4}$\\
CVenII      & $ -7.6^{+0.6}_{-0.5}$& $ 18.9^{+1.3}_{-1.4}$& $159.1^{+10.8}_{-11.6}$& $-24.0^{+43.1}_{-42.9}$& $-455.1^{+54.5}_{-50.9}$& $-77.1^{+6.4}_{-6.8}$\\
CarI        & $-22.2^{+1.6}_{-1.5}$& $-93.1^{+6.7}_{-6.4}$& $-38.0^{+2.7}_{-2.6}$& $-62.8^{+16.7}_{-16.9}$& $-296.9^{+9.8}_{-9.3}$& $143.2^{+22.1}_{-23.4}$\\
CarII       & $ -2.1^{+0.1}_{-0.1}$& $-34.7^{+2.3}_{-2.4}$& $-10.2^{+0.7}_{-0.7}$& $ 87.5^{+9.7}_{-10.1}$& $-547.8^{+6.4}_{-6.8}$& $151.2^{+21.2}_{-20.0}$\\
CarIII      & $ -1.5^{+0.1}_{-0.1}$& $-26.7^{+1.8}_{-1.7}$& $ -7.7^{+0.5}_{-0.5}$& $-40.2^{+7.4}_{-7.2}$& $-396.8^{+9.2}_{-8.5}$& $354.4^{+28.3}_{-30.3}$\\
CBerI       & $ -2.3^{+0.2}_{-0.1}$& $ -4.0^{+0.3}_{-0.3}$& $ 41.5^{+2.7}_{-2.9}$& $240.1^{+17.5}_{-18.4}$& $-261.9^{+18.9}_{-18.4}$& $ 86.5^{+1.4}_{-1.4}$\\
ColI        & $-102.4^{+7.2}_{-7.2}$& $-122.1^{+8.6}_{-8.6}$& $-87.8^{+6.1}_{-6.2}$& $211.4^{+229.7}_{-228.0}$& $-411.7^{+182.4}_{-177.6}$& $  7.5^{+179.3}_{-187.3}$\\
CraI        & $  3.5^{+0.2}_{-0.2}$& $-94.2^{+6.2}_{-6.5}$& $110.1^{+7.6}_{-7.2}$& $ 29.1^{+42.6}_{-41.2}$& $-183.8^{+35.2}_{-33.4}$& $ 38.3^{+30.0}_{-28.6}$\\
CraII       & $ 16.5^{+1.2}_{-1.2}$& $-84.6^{+6.1}_{-5.9}$& $ 79.8^{+5.6}_{-5.8}$& $-52.4^{+19.4}_{-18.8}$& $-137.3^{+13.8}_{-14.2}$& $ -6.0^{+14.2}_{-14.8}$\\
DraI        & $  5.8^{+0.4}_{-0.4}$& $ 64.8^{+4.2}_{-4.5}$& $ 45.1^{+2.9}_{-3.1}$& $ 36.8^{+13.6}_{-14.9}$& $-248.1^{+7.8}_{-7.8}$& $-159.1^{+11.1}_{-11.0}$\\
DraII       & $ -1.7^{+0.1}_{-0.1}$& $ 14.6^{+1.1}_{-1.0}$& $ 13.6^{+1.0}_{-0.9}$& $  8.8^{+5.3}_{-5.5}$& $-158.1^{+7.9}_{-7.5}$& $-340.6^{+8.2}_{-8.5}$\\
EriII       & $-89.2^{+6.5}_{-6.4}$& $-205.7^{+14.9}_{-14.7}$& $-285.3^{+20.7}_{-20.4}$& $-690.5^{+95.6}_{-98.6}$& $107.3^{+76.0}_{-77.0}$& $ 42.4^{+57.0}_{-55.1}$\\
FnxI        & $-34.0^{+2.3}_{-2.2}$& $-45.6^{+3.1}_{-3.0}$& $-126.6^{+8.7}_{-8.2}$& $ 15.6^{+21.9}_{-20.9}$& $-358.9^{+33.0}_{-31.6}$& $ 64.3^{+11.7}_{-12.5}$\\
GruI        & $ 58.0^{+4.1}_{-4.0}$& $-23.8^{+1.6}_{-1.7}$& $-102.0^{+7.0}_{-7.2}$& $ 95.9^{+25.2}_{-26.2}$& $-165.7^{+28.5}_{-29.6}$& $258.1^{+16.8}_{-17.3}$\\
HerI        & $ 95.9^{+6.6}_{-6.4}$& $ 49.3^{+3.4}_{-3.3}$& $ 80.8^{+5.6}_{-5.4}$& $109.6^{+16.0}_{-15.5}$& $-247.0^{+25.7}_{-26.9}$& $ 95.5^{+18.3}_{-17.9}$\\
HorI        & $ -0.4^{+0.0}_{-0.0}$& $-45.5^{+3.1}_{-3.3}$& $-64.6^{+4.4}_{-4.7}$& $-40.4^{+22.0}_{-23.9}$& $-383.9^{+26.5}_{-29.1}$& $132.3^{+20.6}_{-18.6}$\\
HorII       & $ -7.1^{+0.5}_{-0.5}$& $-43.9^{+3.2}_{-3.1}$& $-61.6^{+4.4}_{-4.3}$& $-174.7^{+130.1}_{-118.8}$& $-537.1^{+87.0}_{-92.8}$& $195.0^{+62.7}_{-59.0}$\\
HyaII       & $ 45.6^{+3.1}_{-3.1}$& $-103.5^{+7.0}_{-7.0}$& $ 71.8^{+4.9}_{-4.9}$& $-157.5^{+39.8}_{-42.6}$& $-320.7^{+24.8}_{-26.7}$& $203.2^{+31.9}_{-33.1}$\\
HyiI        & $  9.5^{+0.7}_{-0.7}$& $-19.8^{+1.4}_{-1.4}$& $-16.7^{+1.2}_{-1.2}$& $-167.2^{+13.9}_{-14.5}$& $-428.3^{+26.4}_{-26.9}$& $280.2^{+24.1}_{-23.7}$\\
LeoI        & $-125.7^{+8.3}_{-9.2}$& $-123.3^{+8.2}_{-9.0}$& $203.4^{+14.9}_{-13.5}$& $-159.1^{+37.6}_{-40.8}$& $-288.2^{+37.7}_{-42.0}$& $100.6^{+26.7}_{-31.4}$\\
LeoII       & $-67.8^{+5.0}_{-4.5}$& $-54.2^{+4.0}_{-3.6}$& $206.9^{+13.8}_{-15.2}$& $ 22.6^{+33.8}_{-33.5}$& $-196.2^{+36.6}_{-38.6}$& $ 40.6^{+13.6}_{-14.0}$\\
LeoIV       & $-10.4^{+0.7}_{-0.7}$& $-82.4^{+5.7}_{-5.4}$& $129.9^{+8.5}_{-8.9}$& $-238.1^{+49.6}_{-43.5}$& $-477.0^{+44.9}_{-44.7}$& $-164.6^{+29.9}_{-29.2}$\\
LeoV        & $-15.2^{+1.1}_{-1.1}$& $-89.1^{+6.4}_{-6.4}$& $147.6^{+10.6}_{-10.6}$& $151.9^{+45.5}_{-47.5}$& $-519.5^{+48.6}_{-52.3}$& $-94.6^{+28.5}_{-30.8}$\\
LMC         & $  5.2^{+0.4}_{-0.4}$& $-41.5^{+3.0}_{-2.9}$& $-27.3^{+2.0}_{-1.9}$& $-70.9^{+10.2}_{-9.7}$& $-463.3^{+18.5}_{-17.6}$& $210.8^{+25.1}_{-26.2}$\\
PheII       & $ 31.6^{+2.2}_{-2.1}$& $-23.7^{+1.6}_{-1.7}$& $-69.6^{+4.7}_{-4.9}$& $-46.7^{+42.9}_{-42.6}$& $-467.1^{+61.7}_{-54.3}$& $100.9^{+25.4}_{-27.5}$\\
PhxI        & $ -2.6^{+0.2}_{-0.2}$& $-145.8^{+9.7}_{-10.0}$& $-392.7^{+26.0}_{-26.9}$& $-56.8^{+74.8}_{-81.6}$& $-157.1^{+77.8}_{-78.2}$& $ 81.3^{+29.1}_{-28.8}$\\
PisII       & $ 26.6^{+1.9}_{-1.8}$& $121.7^{+8.9}_{-8.4}$& $-134.1^{+9.3}_{-9.8}$& $292.8^{+59.0}_{-57.4}$& $-477.9^{+45.3}_{-46.5}$& $-66.5^{+39.6}_{-40.5}$\\
RetII       & $ -1.8^{+0.1}_{-0.1}$& $-19.3^{+1.3}_{-1.4}$& $-22.9^{+1.5}_{-1.7}$& $ -6.6^{+5.1}_{-4.8}$& $-338.0^{+20.2}_{-22.7}$& $202.9^{+19.1}_{-17.0}$\\
RetIII      & $  1.3^{+0.1}_{-0.1}$& $-63.8^{+4.4}_{-4.4}$& $-66.3^{+4.6}_{-4.6}$& $205.7^{+267.3}_{-264.3}$& $-124.9^{+165.1}_{-159.5}$& $-256.3^{+153.7}_{-159.5}$\\
SgrI        & $ 25.2^{+1.8}_{-1.7}$& $  2.2^{+0.2}_{-0.1}$& $ -6.2^{+0.4}_{-0.4}$& $217.6^{+6.1}_{-6.0}$& $-270.7^{+19.7}_{-20.4}$& $201.7^{+17.3}_{-16.3}$\\
SgrII       & $ 58.9^{+4.2}_{-3.7}$& $ 18.6^{+1.3}_{-1.2}$& $-26.0^{+1.7}_{-1.8}$& $-21.1^{+20.0}_{-18.3}$& $-358.5^{+40.0}_{-45.2}$& $152.4^{+32.2}_{-31.7}$\\
SclI        & $  1.7^{+0.1}_{-0.1}$& $ -8.0^{+0.5}_{-0.6}$& $-84.4^{+5.5}_{-6.1}$& $  2.4^{+14.1}_{-13.9}$& $-73.4^{+14.1}_{-15.3}$& $-104.9^{+1.5}_{-1.4}$\\
Seg1        & $-11.4^{+0.8}_{-0.8}$& $ -9.2^{+0.6}_{-0.6}$& $ 17.7^{+1.2}_{-1.2}$& $-121.5^{+3.4}_{-3.5}$& $-455.0^{+26.1}_{-25.7}$& $-44.1^{+14.3}_{-14.2}$\\
Seg2        & $-24.0^{+1.7}_{-1.7}$& $ 15.1^{+1.0}_{-1.1}$& $-22.3^{+1.6}_{-1.5}$& $-178.8^{+15.4}_{-15.8}$& $-179.4^{+13.2}_{-13.0}$& $134.9^{+10.5}_{-10.1}$\\
SMC         & $ 23.1^{+1.6}_{-1.6}$& $-38.1^{+2.6}_{-2.6}$& $-46.0^{+3.1}_{-3.2}$& $-13.4^{+9.2}_{-10.0}$& $-439.1^{+25.0}_{-25.2}$& $154.0^{+18.5}_{-18.8}$\\
SxtI        & $-30.7^{+2.2}_{-2.0}$& $-54.6^{+3.9}_{-3.5}$& $ 58.5^{+3.8}_{-4.2}$& $-234.7^{+17.3}_{-15.9}$& $-153.3^{+9.9}_{-10.5}$& $ 62.4^{+11.6}_{-11.4}$\\
TriII       & $-20.8^{+1.4}_{-1.4}$& $ 17.9^{+1.2}_{-1.2}$& $-12.1^{+0.8}_{-0.8}$& $199.8^{+6.7}_{-6.7}$& $-239.8^{+6.1}_{-6.0}$& $248.0^{+9.1}_{-8.8}$\\
TucII       & $ 29.0^{+2.0}_{-2.1}$& $-18.2^{+1.3}_{-1.2}$& $-45.6^{+3.4}_{-3.1}$& $-256.5^{+16.0}_{-14.9}$& $-307.9^{+27.4}_{-24.3}$& $121.0^{+6.1}_{-6.6}$\\
TucIII      & $  9.5^{+0.7}_{-0.7}$& $ -9.6^{+0.7}_{-0.7}$& $-21.0^{+1.5}_{-1.5}$& $ 16.4^{+5.5}_{-5.3}$& $-121.9^{+11.8}_{-11.8}$& $184.6^{+7.4}_{-7.3}$\\
UMaI        & $-52.5^{+3.5}_{-3.7}$& $ 21.3^{+1.5}_{-1.4}$& $ 79.2^{+5.5}_{-5.3}$& $-213.7^{+22.0}_{-22.2}$& $-395.5^{+29.6}_{-31.2}$& $-103.2^{+10.7}_{-10.4}$\\
UMaII       & $-24.2^{+1.5}_{-1.7}$& $ 13.5^{+1.0}_{-0.8}$& $ 21.2^{+1.5}_{-1.3}$& $149.7^{+6.4}_{-6.0}$& $-355.5^{+20.7}_{-22.9}$& $204.7^{+19.7}_{-17.9}$\\
UMiI        & $-12.6^{+0.9}_{-0.9}$& $ 52.6^{+3.6}_{-3.8}$& $ 53.7^{+3.7}_{-3.8}$& $ -9.5^{+5.8}_{-5.8}$& $-197.2^{+4.1}_{-4.1}$& $-156.3^{+3.5}_{-3.5}$\\
Wil1        & $-19.3^{+1.2}_{-1.4}$& $  8.2^{+0.6}_{-0.5}$& $ 32.0^{+2.3}_{-1.9}$& $ 64.8^{+9.8}_{-8.9}$& $-223.5^{+15.1}_{-18.9}$& $ 81.4^{+8.9}_{-7.6}$\\
\hline
 \end{tabular}
 \end{small}
  \end{center}
  \end{table*}

\begin{figure*}
{\begin{center}
    \includegraphics[width=1.05\textwidth,angle=0]{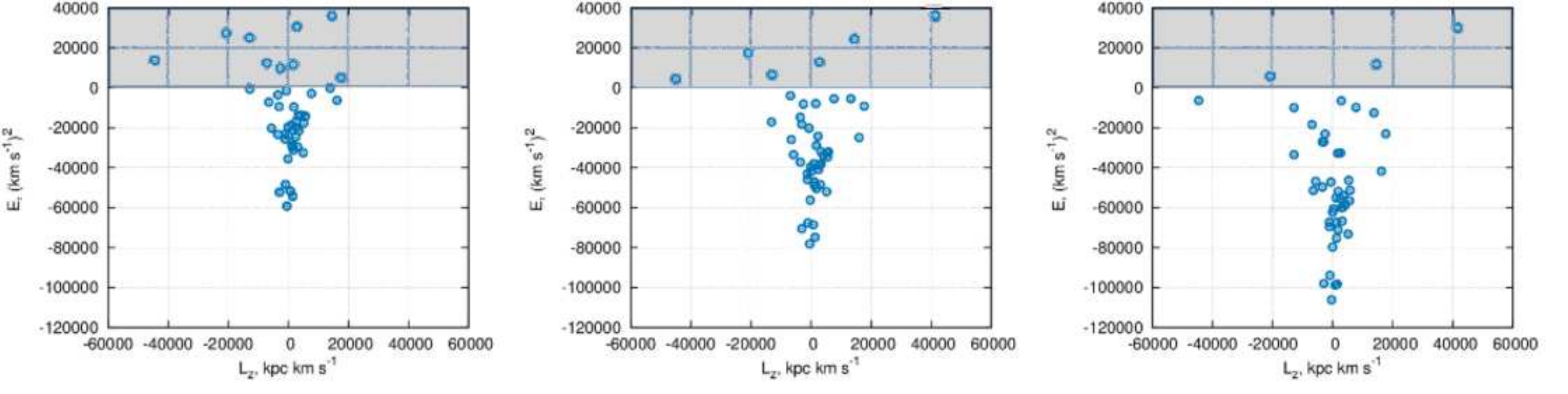}\
    \caption{"$L_Z-E$" diagrams for three models of the Galactic gravitational potential: the NFWBB (left-hand panel), the ASBB (middle panel) and the ASI (right-hand panel). }
\label{f3ELz}
\end{center}}
\end{figure*}

\section{Results}

The orbits of the 47 dwarf galaxies were obtained by integrating Eg. (\ref{eq-motion}) for 13.5 Gyr backward. The orbit properties determined for three models of the Galactic potential (NFWBB, ASBB and ASI) are listed in Table \ref{t:prop}. The comparison of the properties obtained for these models of the Galactic potential with different masses is illustrated in Figs.~\ref{f3ELz} and \ref{fprop}. The orbits of the dwarfs integrated in NFWBB, ASBB and ASI models of the Galactic potential are shown in Figs.~\ref{f11}, \ref{f21} and \ref{f31} respectively. We do not give the orbits of two dwarf galaxies EriII and PisII, which in all three potentials do not show the gravitational connection with the Milky Way. In the Figures we give (X,Y), (X,Z), (Y,Z) projections for each orbit in Cartesian Galactic coordinate system, as well as a radial coordinate of the orbit $r$ as a function of time $T$. The latter gives us an idea of the periodicity of the orbit. To save space, each panel represents the orbits of 5-4 objects, the names of which are given above the pictures. Mostly objects are grouped by proximity of scale.

\subsection{Gravitational Connection of the Dwarf Galaxies with the Milky Way}
\begin{figure*}
{\begin{center}
   \includegraphics[width=0.9\textwidth,angle=0]{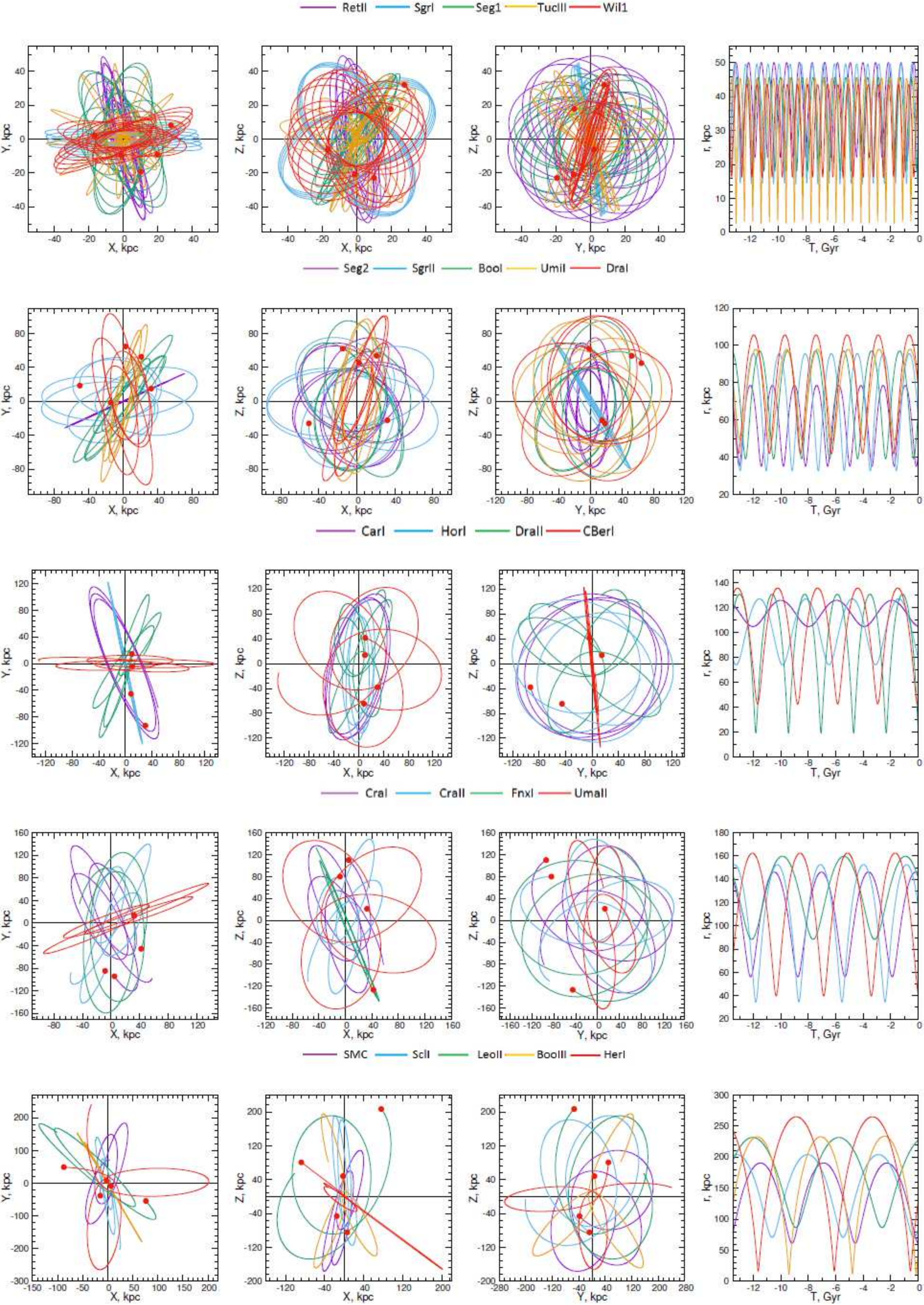}
  \caption{Orbits of the dwarf galaxies in the Galactic potential NFWBB}
\label{f11}
\end{center}}
\end{figure*}

\begin{figure*}
{\begin{center}
   \includegraphics[width=0.9\textwidth,angle=0]{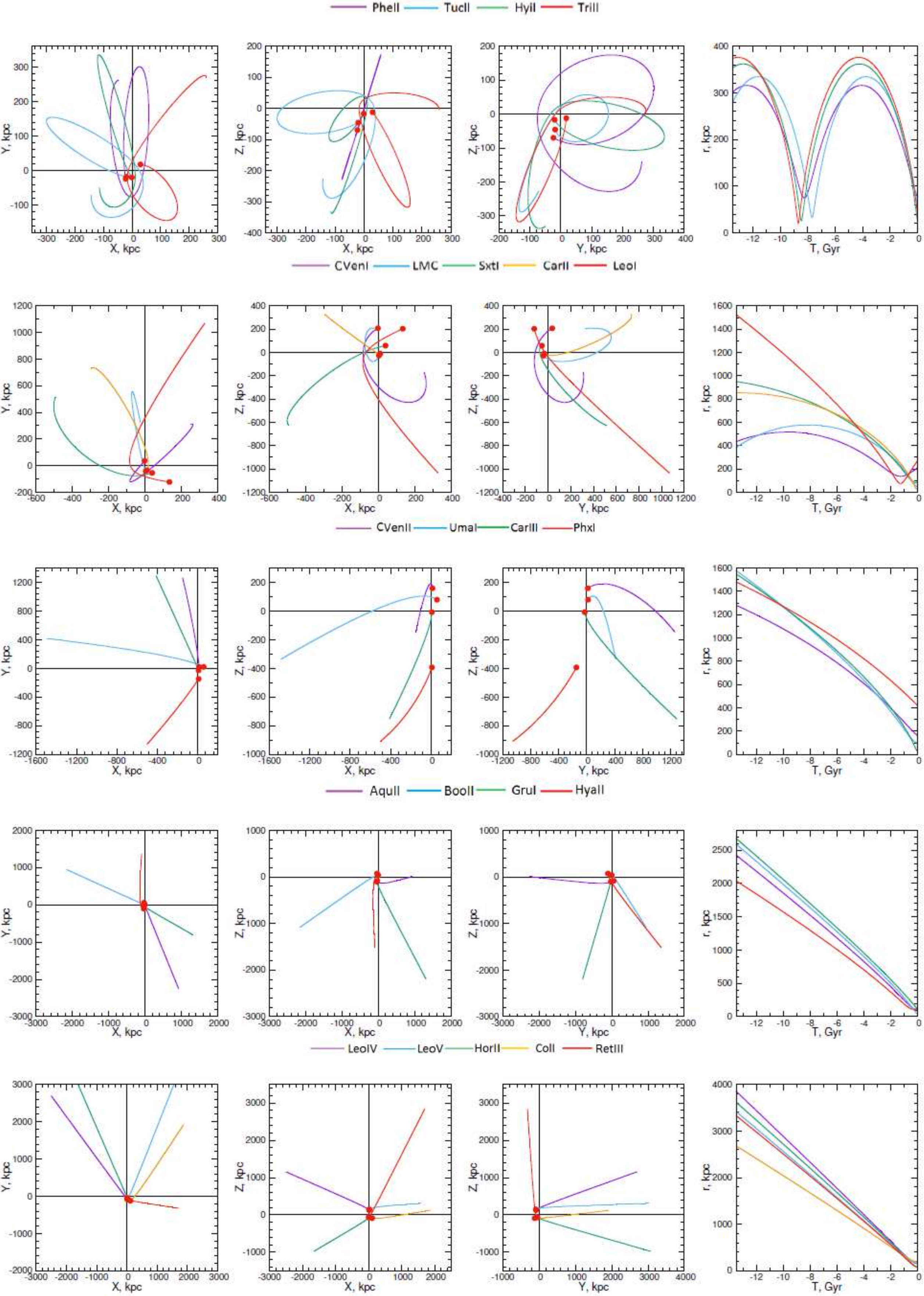}
  \centerline{{\bf Figure~5.} Continued from previous page}
\label{f12}
\end{center}}
\end{figure*}

\begin{figure*}
{\begin{center}
   \includegraphics[width=0.9\textwidth,angle=0]{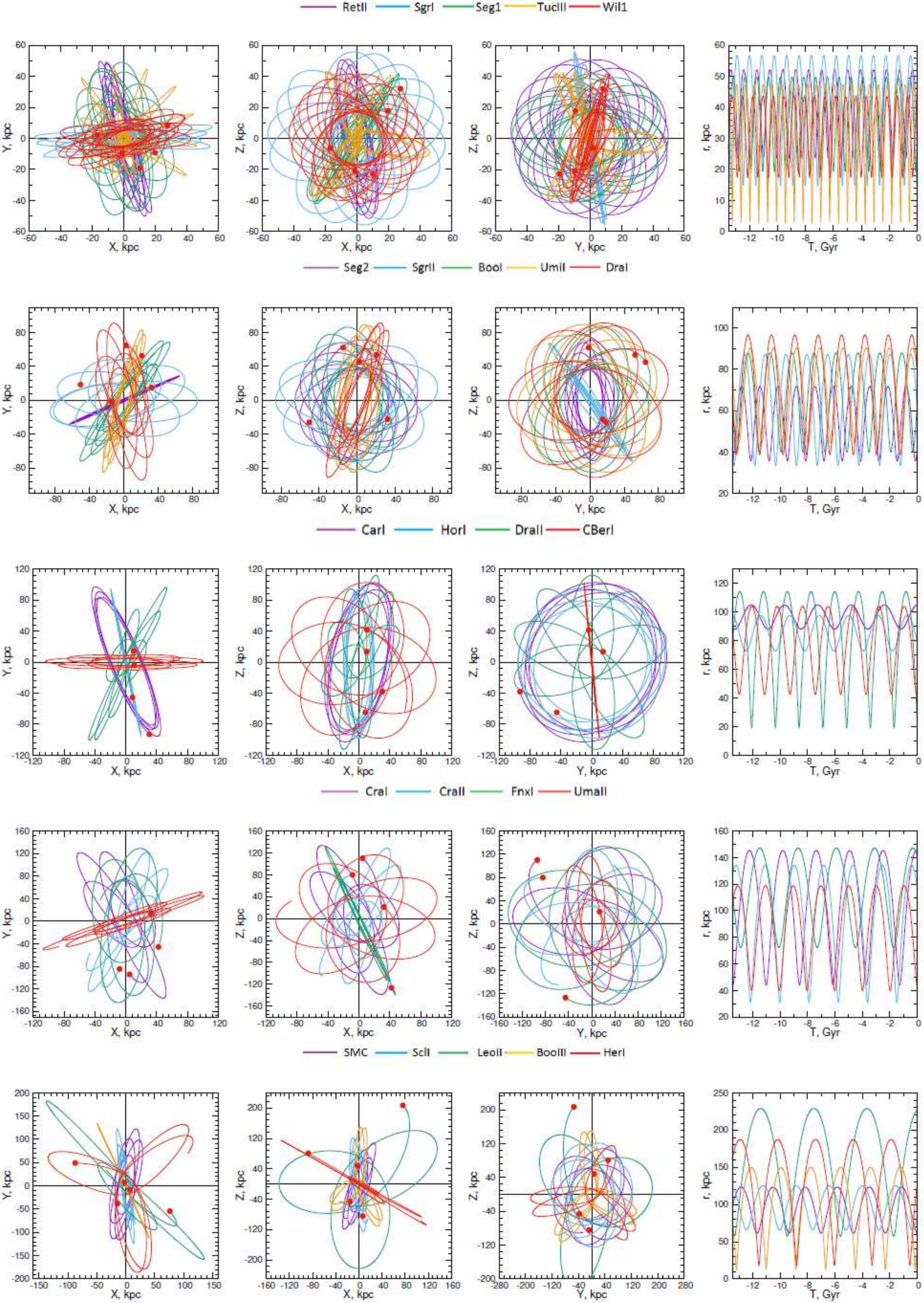}
  \caption{Orbits of the dwarf galaxies in the Galactic potential ASBB}
\label{f21}
\end{center}}
\end{figure*}

\begin{figure*}
{\begin{center}
   \includegraphics[width=0.9\textwidth,angle=0]{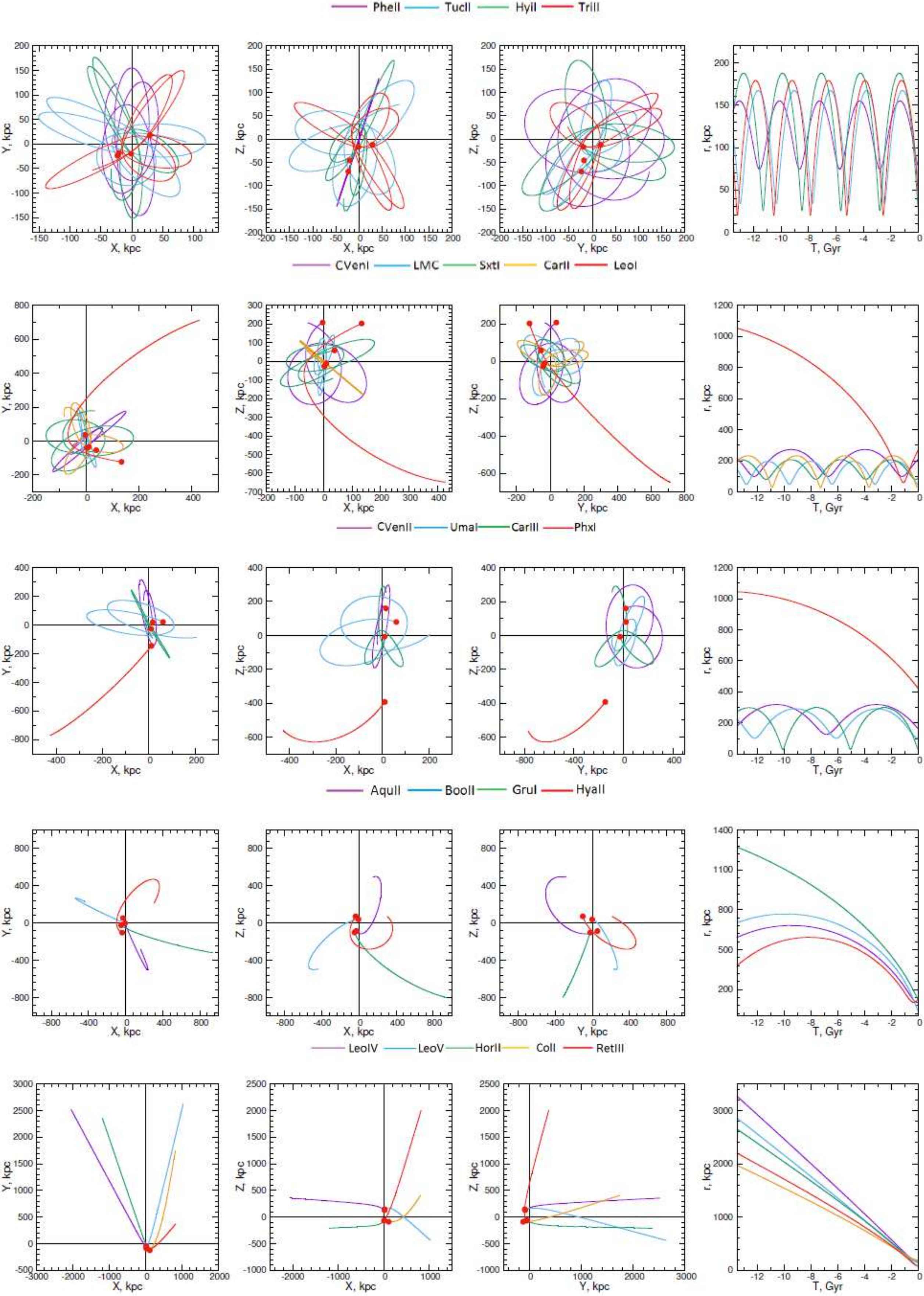}
 \centerline{{\bf Figure~6.} Continued from previous page}
\label{f22}
\end{center}}
\end{figure*}

\begin{figure*}
{\begin{center}
   \includegraphics[width=0.9\textwidth,angle=0]{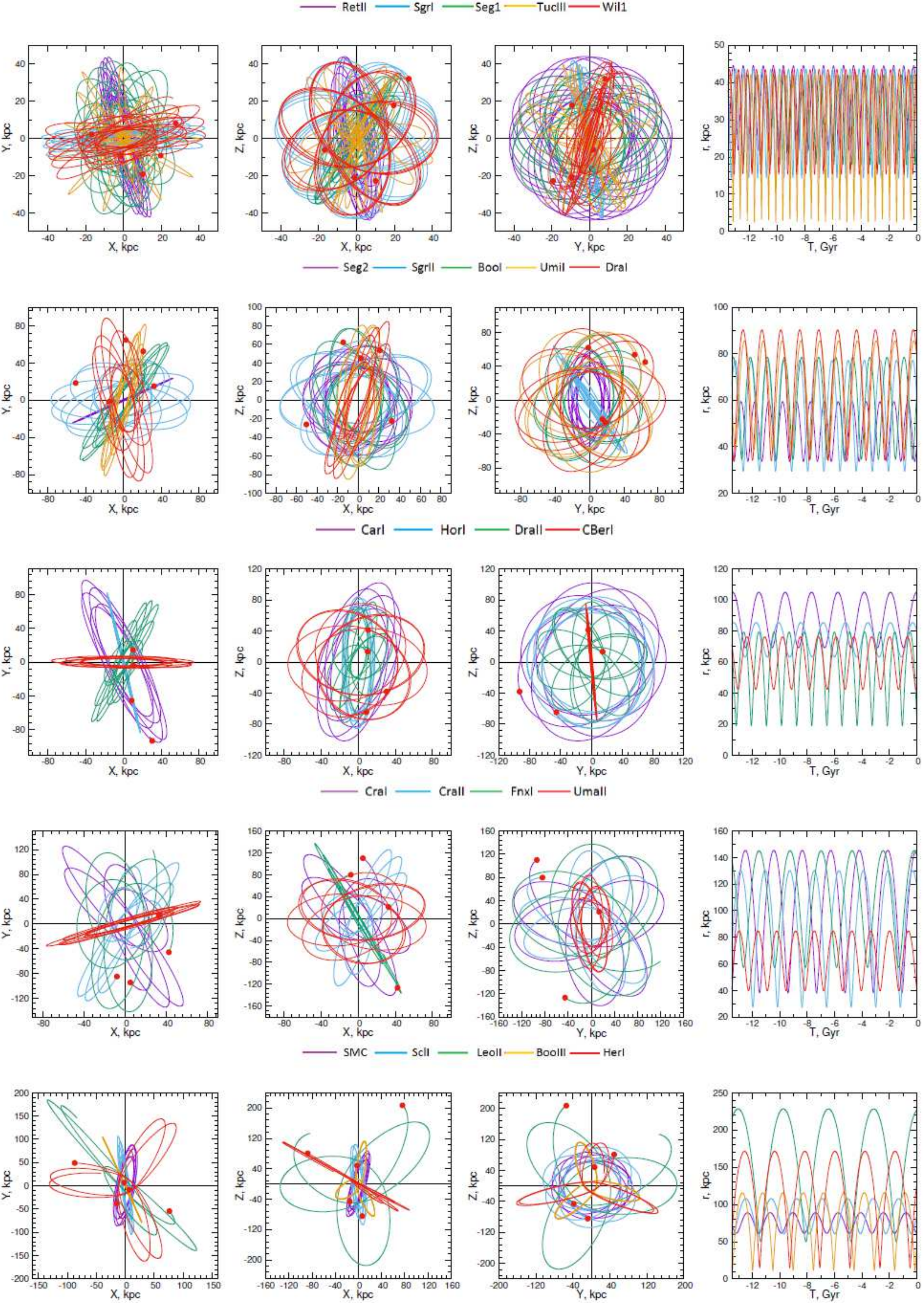}
  \caption{Orbits of the dwarf galaxies in the Galactic potential ASI}
\label{f31}
\end{center}}
\end{figure*}

One of the main questions in the study of the dynamics of dwarf galaxies is their gravitational connection with the Milky Way. Fig.~\ref{f3ELz} shows the diagrams $L_Z-E$, calculated for three models NFWBB, ASBB and ASI of the Galactic potential with different masses $M_{G_{(R \leq 200 kpc)}}=0.75, 1.45$ and $1.9 $ of $10^{12}M_\odot$, respectively. The area of positive energy values $E$ is shown in gray.
The dwarf galaxies that are not gravitationally associated with the Milky Way, having velocities $V_{tot}$ higher than the escape velocities, belong to this area. In the case of the model NFWBB, these are 11 dwarf galaxies (AquII, BooII, ColI, EriII, GruI, HorII, HyaII, LeoIV, LeoV, PisII, RetIII), in the case of the model ASBB, 7 galaxies (ColI, EriII, HorII, LeoIV, LeoV, PisII, RetIII), and in the case of the model ASI, 4 galaxies (EriII, LeoIV, LeoV, PisII).  Note that in Table~\ref{t:prop} for such dwarfs the orbit parameters apo, peri, ecc, $\theta$ and $T_r$ are not determined. There are also several dwarf galaxies with a very weak gravitational connection with the Milky Way (CVenII, CarIII, SxtI, UmaI in NFWBB model, LeoI both in NFWBB and ASBB models, PhxI in all three models). There are several dwarf galaxies with a very weak gravitational connection with the Milky Way. Although they are characterized by negative total energy $E$, it is not enough for reliable determination of orbital parameters. Table~\ref{t:prop} and Figs.~\ref{f11}--\ref{f31}  allow us to trace the dependence of the orbital motion of the dwarf galaxies on the model of the Galactic gravitational potential and mass of the Galaxy. Thus, the mass of the Galaxy determines degree of gravitational connection with dwarf galaxies. Therefore, further work on refining the rotation curve of our Galaxy and on this basis refining a model of its potential and mass is of great importance.

\subsection{The Comparison of the Orbits Integrated in the Three Models of the Galactic Potential}
\begin{figure*}
{\begin{center}
          \includegraphics[width=0.22\textwidth,angle=-90]{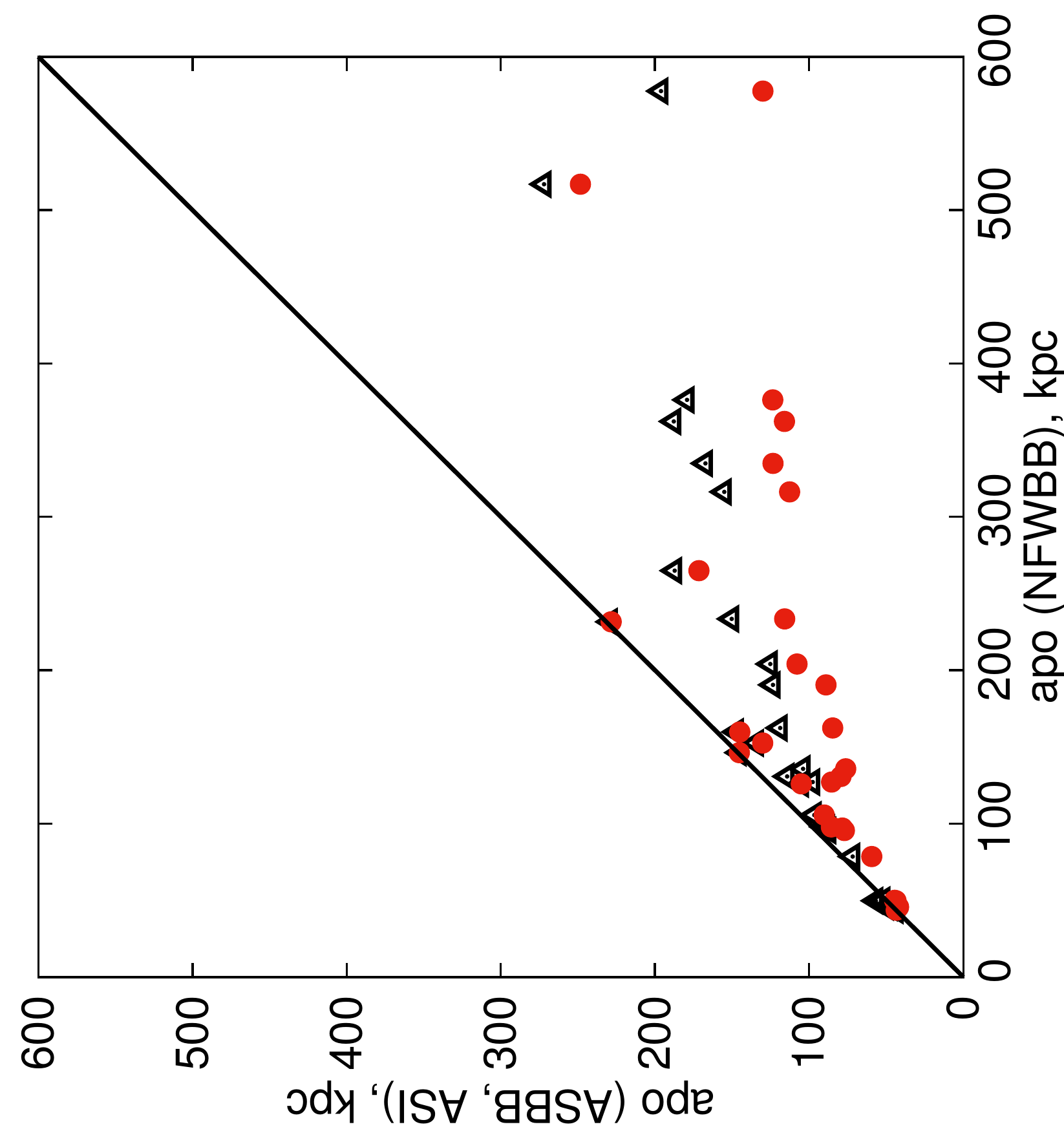}
     \includegraphics[width=0.22\textwidth,angle=-90]{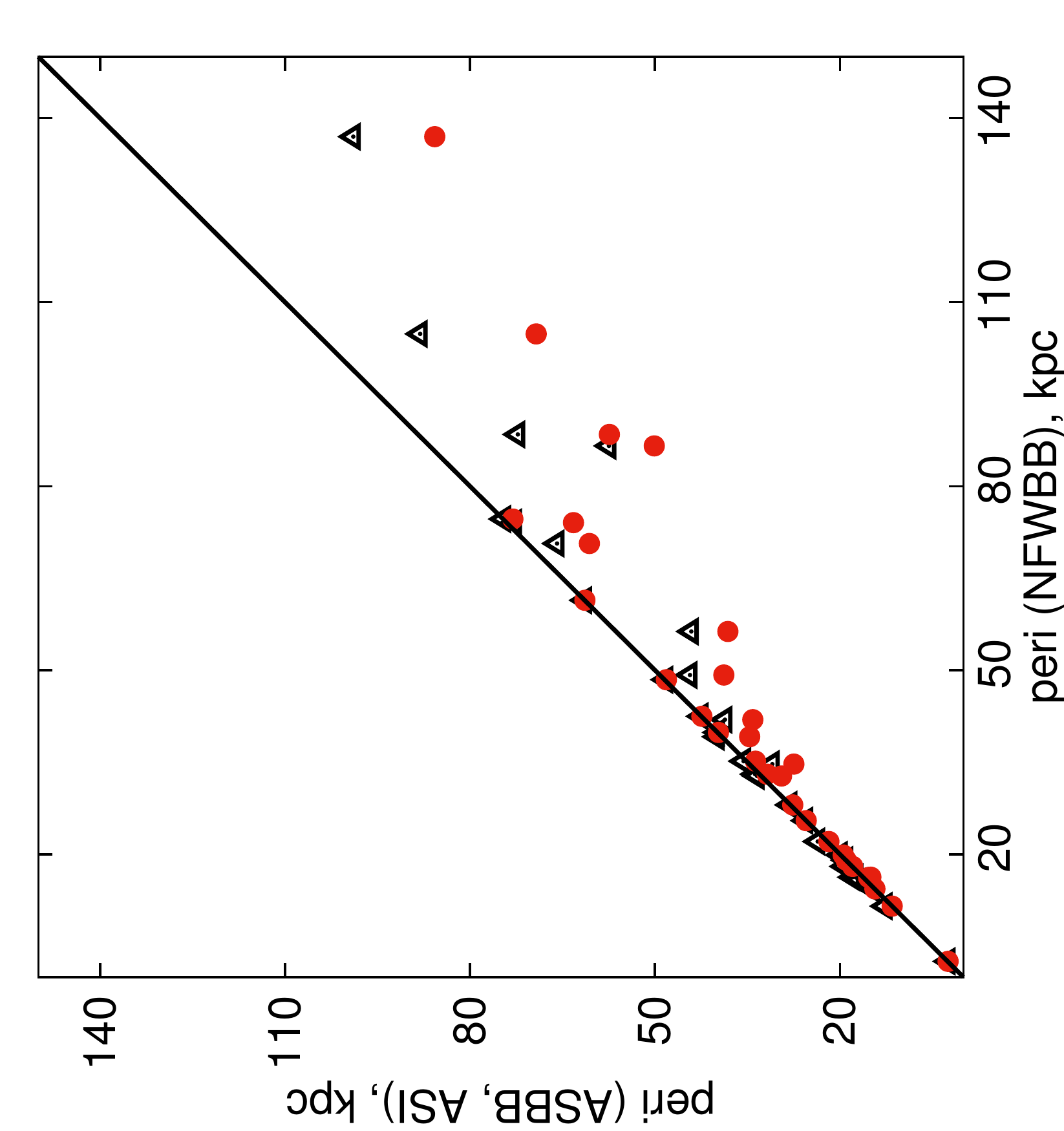}
   \includegraphics[width=0.22\textwidth,angle=-90]{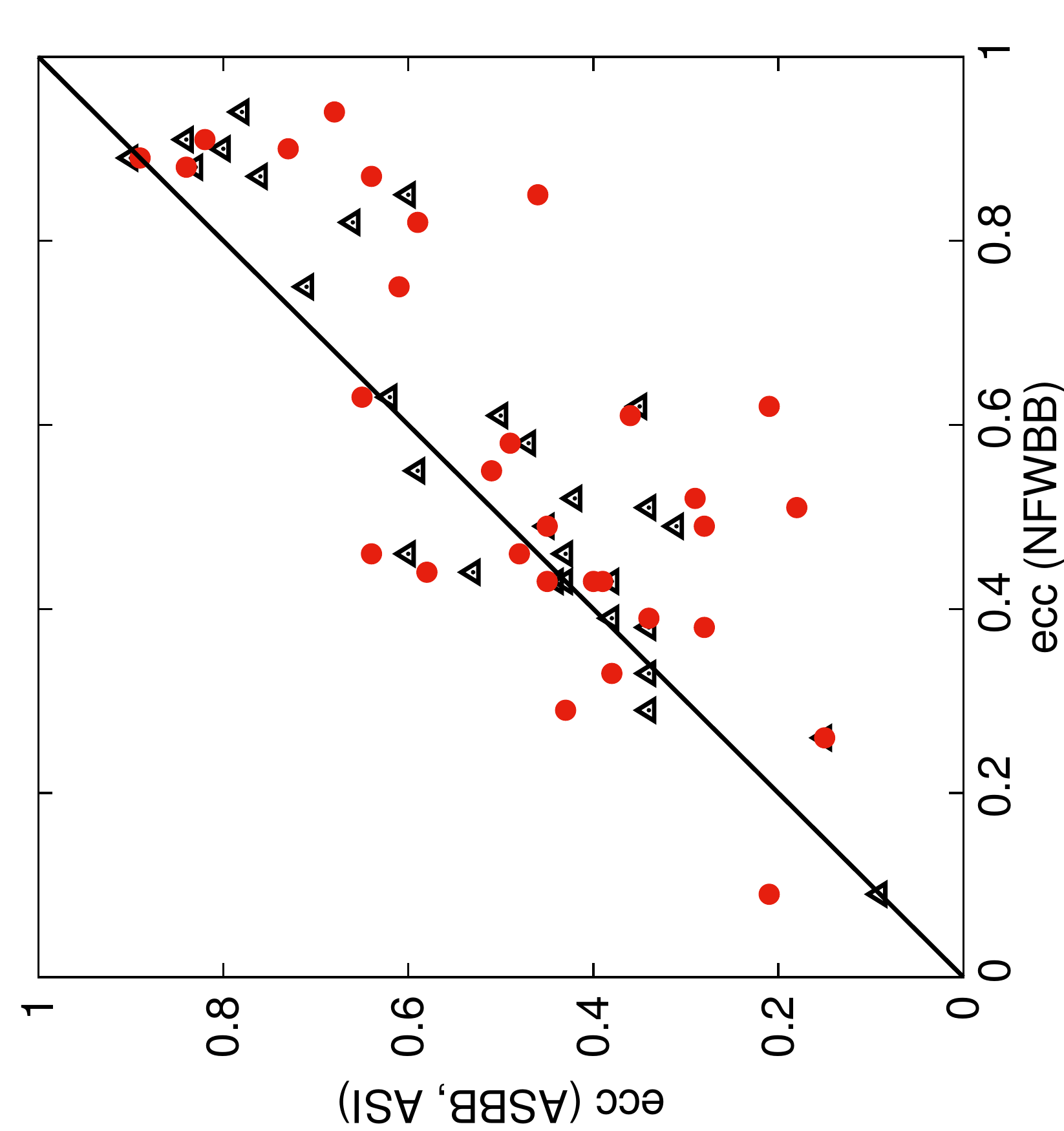}\

\medskip
      \includegraphics[width=0.22\textwidth,angle=-90]{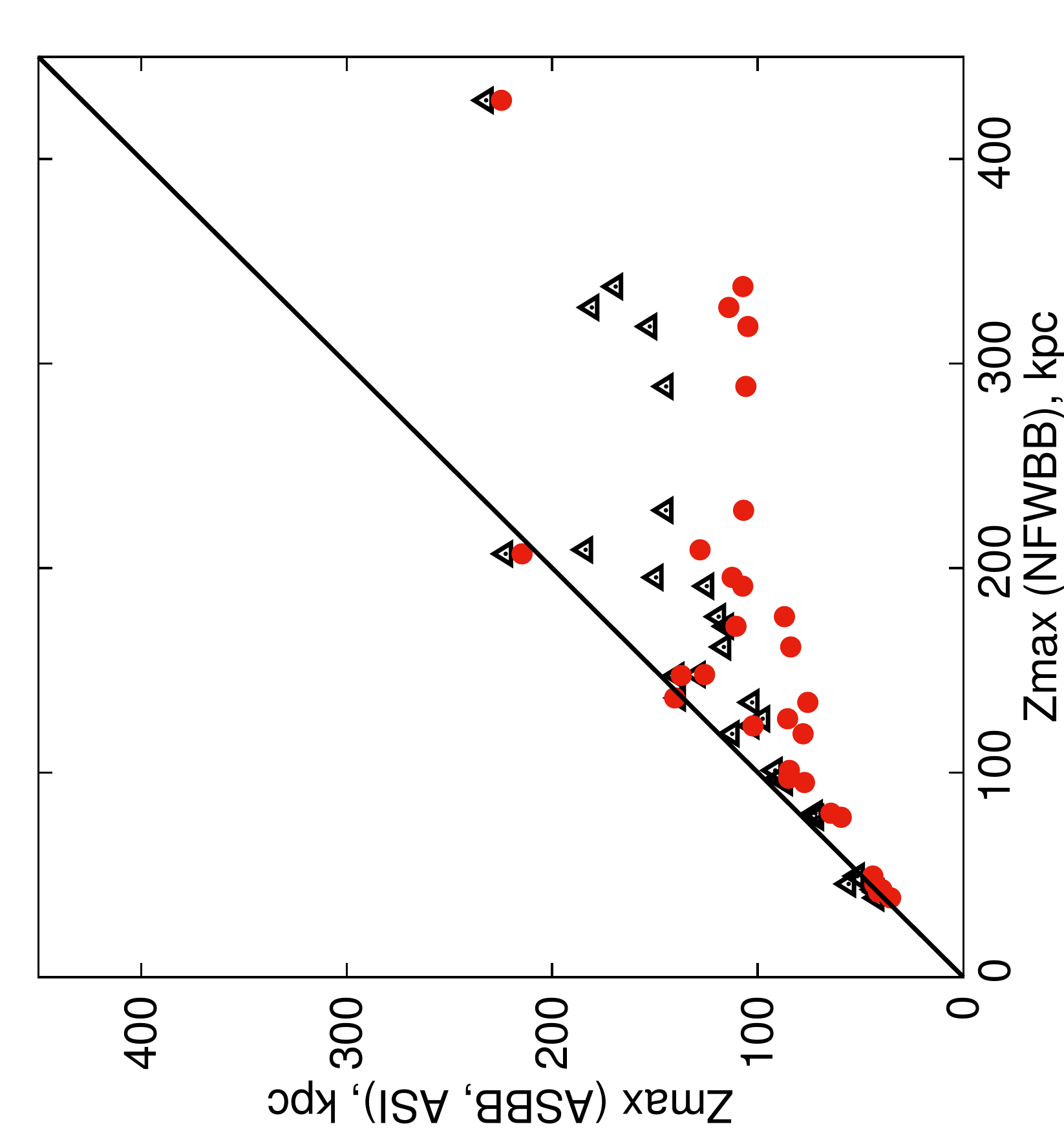}
     \includegraphics[width=0.22\textwidth,angle=-90]{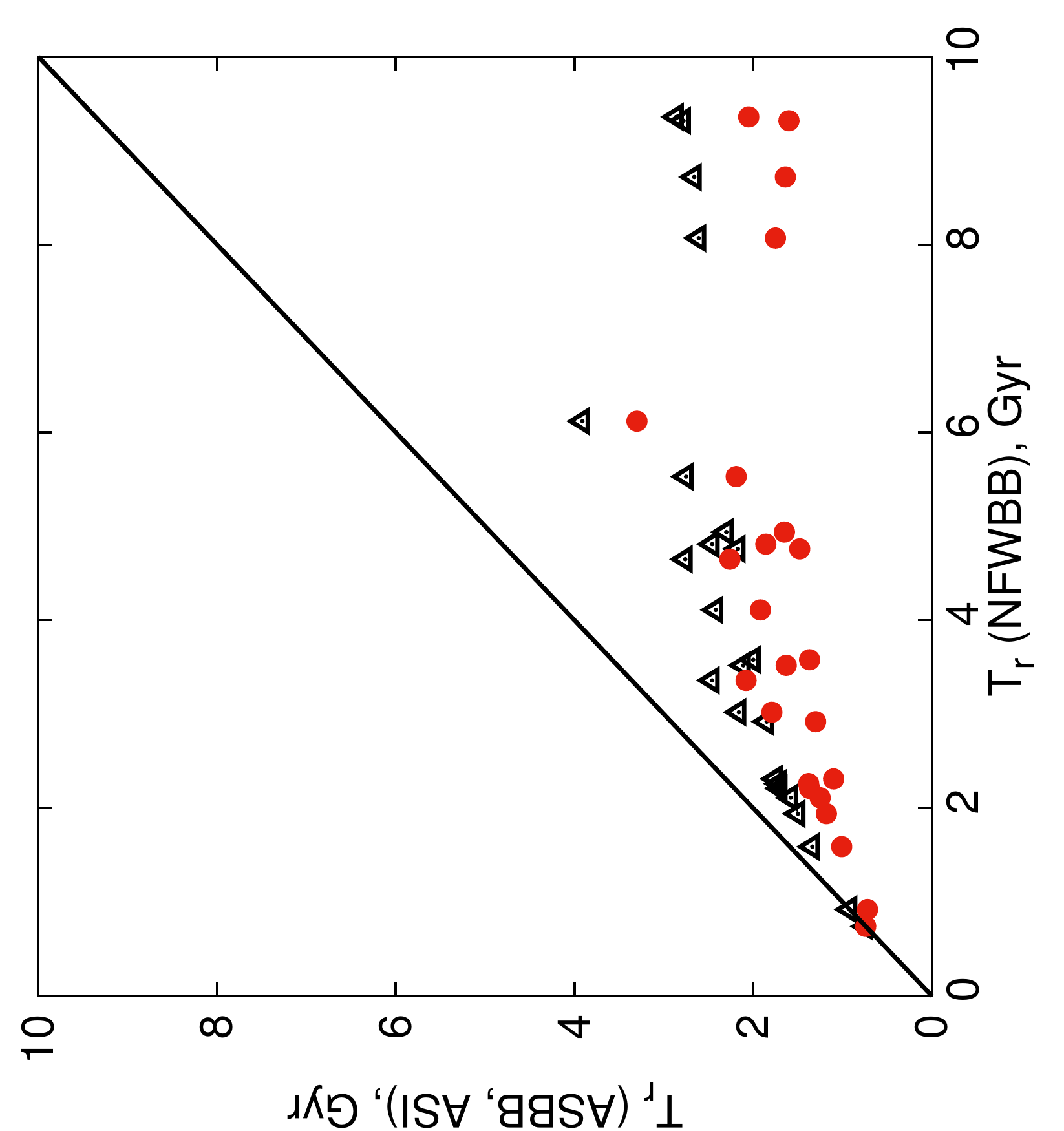}
   \includegraphics[width=0.22\textwidth,angle=-90]{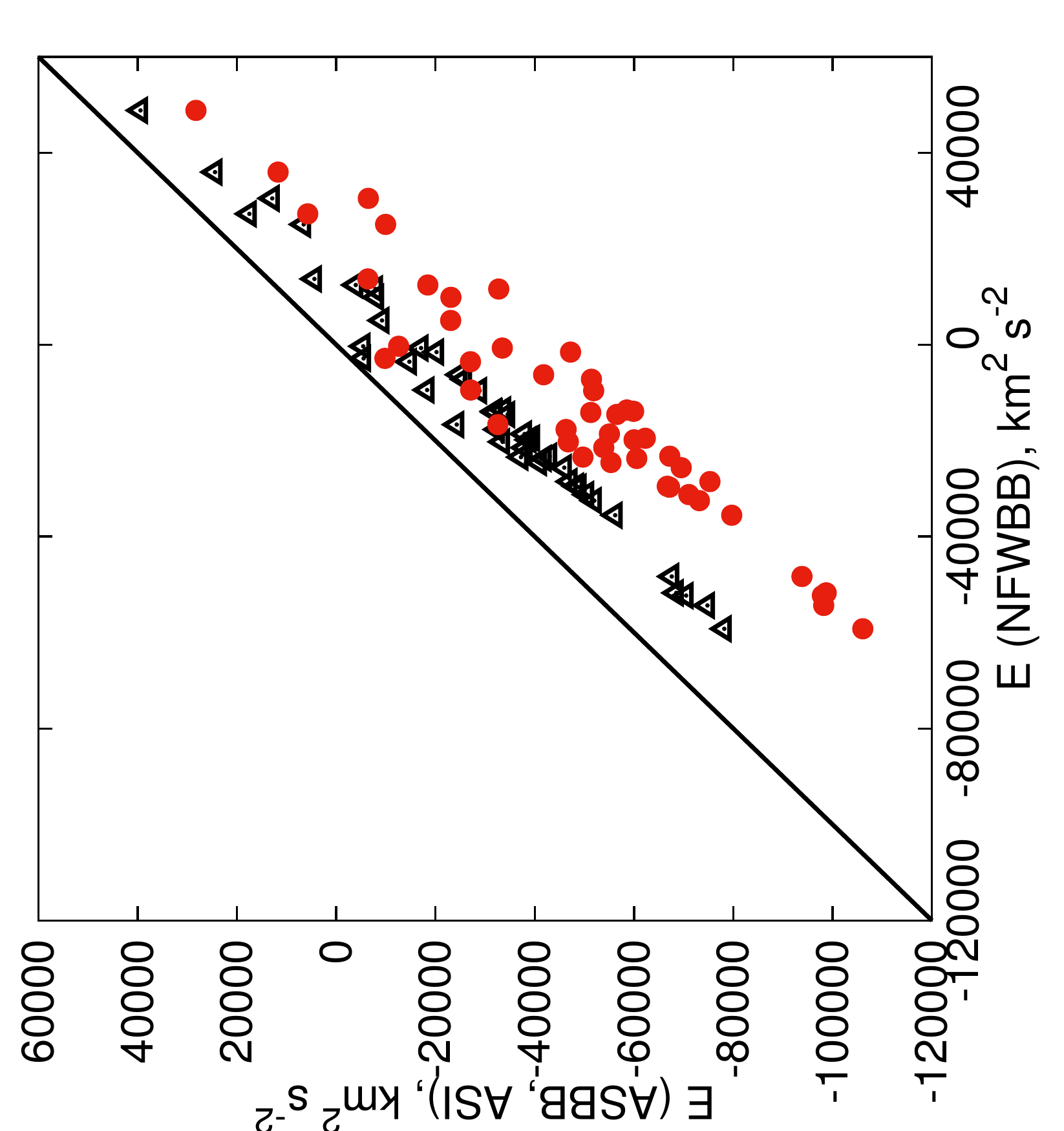}

\caption{Comparison of the orbital parameters (apo, peri, ecc, $Z_{max}$, $T_r$ and $E$) obtained in NFWBB, ASBB and ASI axisymmetric potentials. The parameters obtained in the ASBB (black triangles) and the ASI (red dots) models of the Galactic potential are compared with corresponding orbit parameters obtained in the NFWBB model. In each panel we plot the line of coincidence.}
\label{fprop}
\end{center}}
\end{figure*}

In Fig.~\ref{fprop} we present a comparison of the orbital parameters of the dwarf galaxies obtained using three models (NFWBB, ASBB, ASI) of the Galactic potential. These parameters are apo, peri, ecc, $Z_{max}$,
$T_r$ and $E$. Other orbital parameters listed in the Table~\ref{t:prop} ($\Pi, \Theta, V_{tot}, \theta, L_Z$) depend slightly on the model of the Galactic potential, therefore, we do not give graphs of their comparison.

In Fig.~\ref{fprop} the parameters obtained in the ASBB (black triangles) and the ASI (red dots) models of the Galactic potential are compared with corresponding orbit parameters obtained in the NFWBB model.In each panel the line of coincidence is plotted. It can be seen from the pictures that with an increase in the mass of the Galaxy, at large distances, where the rotation curves of the Galaxy corresponding to the different  models of the potential diverge significantly, the parameters apo, peri, $Z_{max}$, $T_r$ tend to decrease in magnitude. Almost all points in the pictures are below the coincidence line. The behavior of the eccentricity, which depends on both apocentric and pericentric distance, is different for different objects. For majority of the dwarfs the eccentricity decreases with an increase of the mass of the Galaxy. That is, a larger mass of the Galaxy leads to a more circular orbit.  The total energy of the dwarfs with an increase in the mass of the Galaxy proportionally shifts to a more negative zone, parallel to the coincidence line, since the fraction of potential energy, which is negative, increases.

It is currently difficult to determine the only valid Galaxy model. This problem requires further refinement of the model of the Galaxy and its mass. Further refinement of the three-dimensional dynamics of galactic objects also requires further refinement of astrometric data.

\begin{table*}
 \begin{center}
 \caption[]
 {\small\baselineskip=1.0ex
Dwarf orbital properties. For each dwarf
we quote values derived from orbits integrated for 13.5 Gyr backward in three potentials of Models NFWBB,
ASBB and ASI in the first, second, and third row,
respectively.  }
 \label{t:prop}
 \begin{scriptsize}
 \begin{tabular}{|l|r|r|r|r|r|r|r|r|r|r|r|}\hline
 Name &$d_{GC}$&$\Pi$ &$\Theta$ &$V_{tot}$ &apo& peri &ecc&incl. $\theta$&$T_r$&$L_Z$&$E$   \\
    &[kpc]&[km/s]&[km/s]&[km/s]& [kpc]& [kpc]&  &[deg]&[Gyr]&[kpc km/s]&[(km/s)$^2$]\\\hline
AquII     &     105  & $  266^{+   30}_{-   31}$& $  -43^{+   26}_{-   31}$& $  299^{+   30}_{-   28}$& --- & --- & --- & --- & ---  & $  -2629 $& $      9873$\\
  &     106  & $  261^{+ 26}_{- 26}$& $  -41^{+ 40}_{- 39}$& $  295^{+ 29}_{- 20}$& $  682^{+  0}_{-725}$& $  103^{+   17}_{-   25}$& $0.74^{+0.26}_{-0.13}$& $94.7^{+ 4.5}_{- 4.2}$& $18.70^{+0.00}_{-20.14}$& $  -2483 $& $     -8157$\\
  &     106  & $  264^{+ 25}_{- 31}$& $  -42^{+ 32}_{- 28}$& $  297^{+ 27}_{- 27}$& $  280^{+201}_{-125}$& $  103^{+    6}_{-    7}$& $0.46^{+0.20}_{-0.14}$& $94.7^{+ 3.6}_{- 3.3}$& $5.04^{+5.63}_{-3.01}$& $  -2548 $& $    -23102$\\
BooI      &      64  & $  126^{+ 13}_{- 11}$& $  118^{+ 30}_{- 22}$& $  185^{+ 24}_{- 16}$& $   97^{+ 30}_{- 14}$& $   39^{+    9}_{-    8}$& $0.43^{+0.06}_{-0.02}$& $80.2^{+ 2.0}_{- 1.7}$& $2.04^{+0.78}_{-0.40}$& $   1741 $& $    -31258$\\
  &      64  & $  127^{+ 12}_{- 13}$& $  122^{+ 24}_{- 29}$& $  188^{+ 20}_{- 22}$& $   88^{+ 16}_{- 10}$& $   40^{+    8}_{-    8}$& $0.38^{+0.06}_{-0.03}$& $80.0^{+ 1.4}_{- 2.0}$& $1.60^{+0.30}_{-0.23}$& $   1814 $& $    -50438$\\
  &      64  & $  126^{+ 11}_{- 11}$& $  119^{+ 29}_{- 29}$& $  186^{+ 22}_{- 18}$& $   78^{+  8}_{-  5}$& $   35^{+    8}_{-    7}$& $0.39^{+0.07}_{-0.06}$& $80.2^{+ 1.7}_{- 2.1}$& $1.25^{+0.16}_{-0.13}$& $   1756 $& $    -71036$\\
BooII     &      39  & $ -307^{+   30}_{-   25}$& $  244^{+   40}_{-   30}$& $  392^{+   30}_{-   25}$& --- & --- & --- & --- & ---  & $   1637 $& $     11583$\\
  &      39  & $ -305^{+   33}_{-   29}$& $  248^{+   28}_{-   29}$& $  394^{+   26}_{-   26}$& --- & --- & --- & --- & ---  & $   1669 $& $     -7932$\\
  &      39  & $ -305^{+ 36}_{- 26}$& $  247^{+ 33}_{- 33}$& $  393^{+ 23}_{- 27}$& $  234^{+143}_{- 88}$& $   39^{+    5}_{-    7}$& $0.72^{+0.08}_{-0.09}$& $83.8^{+ 0.8}_{- 1.0}$& $3.29^{+3.41}_{-1.85}$& $   1635 $& $    -32754$\\
BooIII    &      49  & $  -66^{+ 26}_{- 40}$& $   -7^{+ 24}_{- 21}$& $  274^{+ 16}_{-  5}$& $  234^{+144}_{-121}$& $   12^{+    6}_{-    3}$& $0.91^{+0.05}_{-0.04}$& $90.5^{+ 2.1}_{- 2.0}$& $4.64^{+4.18}_{-2.89}$& $    -53 $& $    -19518$\\
  &      49  & $  -71^{+ 25}_{- 30}$& $   -6^{+ 22}_{- 26}$& $  275^{+ 17}_{-  6}$& $  150^{+ 86}_{- 60}$& $   13^{+    5}_{-    3}$& $0.84^{+0.07}_{-0.03}$& $90.5^{+ 1.7}_{- 1.9}$& $2.18^{+0.79}_{-0.65}$& $    -42 $& $    -39549$\\
  &      49  & $  -68^{+ 25}_{- 33}$& $   -7^{+ 24}_{- 23}$& $  274^{+ 16}_{-  7}$& $  116^{+ 70}_{- 51}$& $   12^{+    4}_{-    3}$& $0.82^{+0.07}_{-0.05}$& $90.5^{+ 1.9}_{- 2.0}$& $1.48^{+0.32}_{-0.31}$& $    -55 $& $    -62243$\\
CVenI     &     212  & $  110^{+ 28}_{- 38}$& $  -84^{+ 35}_{- 28}$& $  154^{+ 26}_{- 28}$& $  517^{+  0}_{-442}$& $  137^{+   28}_{-   44}$& $0.58^{+0.33}_{-0.08}$& $96.4^{+ 2.2}_{- 2.1}$& $16.82^{+0.00}_{-13.69}$& $  -3022 $& $     -9436$\\
  &     212  & $  105^{+ 41}_{- 27}$& $  -84^{+ 32}_{- 34}$& $  150^{+ 41}_{- 21}$& $  272^{+139}_{- 93}$& $   99^{+   47}_{-   25}$& $0.47^{+0.20}_{-0.08}$& $96.6^{+ 2.3}_{- 1.5}$& $5.50^{+5.01}_{-2.42}$& $  -3008 $& $    -18295$\\
  &     209  & $  108^{+ 32}_{- 36}$& $  -84^{+ 34}_{- 34}$& $  153^{+ 34}_{- 27}$& $  248^{+ 73}_{- 45}$& $   86^{+   37}_{-   30}$& $0.49^{+0.12}_{-0.09}$& $96.4^{+ 1.9}_{- 2.0}$& $4.09^{+2.64}_{-1.52}$& $  -3005 $& $    -27067$\\
CVenII    &     163  & $ -144^{+   46}_{-   47}$& $ -138^{+   46}_{-   40}$& $  211^{+   51}_{-   47}$& --- & --- & --- & --- & ---  & $  -3416 $& $     -3575$\\
  &     161  & $ -147^{+ 46}_{- 48}$& $ -141^{+ 46}_{- 53}$& $  216^{+ 58}_{- 44}$& $  319^{+ 77}_{-282}$& $  126^{+    0}_{-  113}$& $0.43^{+0.49}_{-0.08}$& $96.4^{+ 1.6}_{- 1.4}$& $7.45^{+2.39}_{-7.26}$& $  -3494 $& $    -14728$\\
  &     162  & $ -145^{+ 50}_{- 49}$& $ -140^{+ 57}_{- 35}$& $  213^{+ 48}_{- 53}$& $  222^{+125}_{-123}$& $  114^{+   15}_{-   70}$& $0.32^{+0.38}_{-0.05}$& $96.4^{+ 1.8}_{- 1.0}$& $3.92^{+3.54}_{-2.96}$& $  -3466 $& $    -27017$\\
CarI      &     104  & $   55^{+  9}_{-  9}$& $   36^{+ 19}_{- 12}$& $  164^{+ 20}_{- 21}$& $  126^{+ 97}_{- 54}$& $  105^{+    8}_{-   32}$& $0.09^{+0.26}_{-0.07}$& $78.1^{+ 6.4}_{- 3.7}$& $4.05^{+2.72}_{-1.91}$& $   3568 $& $    -21443$\\
  &     104  & $   59^{+ 13}_{-  8}$& $   35^{+ 15}_{- 17}$& $  165^{+ 22}_{- 16}$& $  105^{+ 15}_{-  6}$& $   88^{+   14}_{-   17}$& $0.09^{+0.11}_{-0.04}$& $78.7^{+ 4.6}_{- 5.5}$& $2.43^{+0.34}_{-0.26}$& $   3418 $& $    -38049$\\
  &     106  & $   57^{+  8}_{- 11}$& $   36^{+ 14}_{- 18}$& $  165^{+ 18}_{- 20}$& $  105^{+  8}_{-  5}$& $   69^{+   14}_{-   14}$& $0.21^{+0.10}_{-0.09}$& $78.3^{+ 5.1}_{- 6.0}$& $1.92^{+0.20}_{-0.18}$& $   3506 $& $    -53875$\\
CarII     &      38  & $  251^{+  8}_{-  8}$& $ -178^{+  8}_{- 13}$& $  346^{+ 12}_{- 10}$& $  853^{+  0}_{-878}$& $   28^{+    2}_{-    2}$& $0.94^{+0.06}_{-0.03}$& $127.3^{+ 2.4}_{- 3.3}$& $26.25^{+0.00}_{-27.14}$& $  -6451 $& $     -7199$\\
  &      37  & $  256^{+  7}_{-  9}$& $ -180^{+ 11}_{- 11}$& $  350^{+ 13}_{-  9}$& $  232^{+ 59}_{- 35}$& $   28^{+    2}_{-    2}$& $0.78^{+0.03}_{-0.02}$& $127.3^{+ 3.1}_{- 2.3}$& $3.60^{+1.30}_{-0.67}$& $  -6502 $& $    -25884$\\
  &      38  & $  252^{+  7}_{-  8}$& $ -179^{+ 10}_{- 10}$& $  348^{+ 12}_{- 11}$& $  145^{+ 21}_{- 14}$& $   28^{+    2}_{-    2}$& $0.68^{+0.02}_{-0.02}$& $127.4^{+ 2.8}_{- 3.2}$& $1.97^{+0.27}_{-0.18}$& $  -6500 $& $    -51419$\\
CarIII    &      29  & $  142^{+    7}_{-   11}$& $  -21^{+    8}_{-    8}$& $  389^{+   33}_{-   23}$& --- & --- & --- & --- & ---  & $   -602 $& $     -1531$\\
  &      29  & $  147^{+  8}_{- 10}$& $  -23^{+  8}_{-  8}$& $  391^{+ 29}_{- 27}$& $  298^{+135}_{-217}$& $   29^{+    1}_{-    2}$& $0.82^{+0.11}_{-0.08}$& $93.3^{+ 1.1}_{- 1.2}$& $5.04^{+4.01}_{-4.60}$& $   -650 $& $    -20224$\\
  &      29  & $  144^{+  9}_{-  7}$& $  -22^{+  9}_{-  8}$& $  390^{+ 29}_{- 30}$& $  161^{+ 90}_{- 62}$& $   29^{+    2}_{-    2}$& $0.69^{+0.08}_{-0.09}$& $93.2^{+ 1.3}_{- 1.1}$& $2.19^{+1.66}_{-1.09}$& $   -635 $& $    -47186$\\
CBerI     &      43  & $ -233^{+ 16}_{- 18}$& $  -95^{+ 18}_{- 21}$& $  268^{+ 19}_{- 15}$& $  136^{+ 87}_{- 45}$& $   43^{+    3}_{-    2}$& $0.52^{+0.13}_{-0.09}$& $95.4^{+ 1.0}_{- 1.1}$& $2.91^{+2.25}_{-1.12}$& $  -1075 $& $    -25626$\\
  &      43  & $ -231^{+ 18}_{- 19}$& $  -99^{+ 19}_{- 20}$& $  268^{+ 18}_{- 15}$& $  104^{+ 27}_{- 18}$& $   43^{+    3}_{-    2}$& $0.42^{+0.08}_{-0.06}$& $95.6^{+ 1.1}_{- 1.1}$& $1.85^{+0.39}_{-0.27}$& $  -1127 $& $    -45916$\\
  &      43  & $ -232^{+ 13}_{- 20}$& $  -96^{+ 19}_{- 20}$& $  268^{+ 19}_{- 12}$& $   76^{+ 17}_{- 10}$& $   42^{+    2}_{-    3}$& $0.29^{+0.07}_{-0.05}$& $95.5^{+ 1.1}_{- 1.0}$& $1.30^{+0.22}_{-0.14}$& $  -1096 $& $    -69489$\\
ColI      &     188  & $  -34^{+  193}_{-  167}$& $ -269^{+  214}_{-  153}$& $  272^{+  227}_{-   44}$& --- & --- & --- & --- & ---  & $ -44403 $& $     13687$\\
  &     188  & $  -31^{+  200}_{-  199}$& $ -273^{+  186}_{-  234}$& $  275^{+  315}_{-   28}$& --- & --- & --- & --- & ---  & $ -44946 $& $      4369$\\
  &     187  & $  -33^{+  189}_{-  185}$& $ -271^{+  175}_{-  217}$& $  273^{+  291}_{-   26}$& --- & --- & --- & --- & ---  & $ -44629 $& $     -6363$\\
CraI      &     144  & $  -74^{+ 37}_{- 34}$& $  -36^{+ 42}_{- 44}$& $   94^{+ 45}_{- 22}$& $  146^{+ 57}_{- 22}$& $   56^{+   50}_{-   24}$& $0.44^{+0.21}_{-0.21}$& $104.7^{+16.3}_{-14.5}$& $3.43^{+2.45}_{-0.86}$& $  -3436 $& $    -23427$\\
  &     145  & $  -69^{+ 31}_{- 39}$& $  -37^{+ 51}_{- 39}$& $   91^{+ 46}_{- 18}$& $  145^{+ 14}_{-  6}$& $   44^{+   36}_{-   19}$& $0.53^{+0.17}_{-0.22}$& $105.3^{+21.2}_{-12.3}$& $2.43^{+0.52}_{-0.19}$& $  -3460 $& $    -37226$\\
  &     145  & $  -72^{+ 31}_{- 34}$& $  -36^{+ 44}_{- 41}$& $   93^{+ 42}_{- 10}$& $  145^{+ 13}_{-  7}$& $   38^{+   28}_{-   11}$& $0.58^{+0.10}_{-0.19}$& $104.9^{+16.8}_{-13.1}$& $2.08^{+0.35}_{-0.11}$& $  -3438 $& $    -49697$\\
CraII     &     117  & $ -122^{+ 14}_{- 15}$& $   30^{+ 21}_{- 19}$& $  126^{+ 18}_{- 11}$& $  153^{+ 28}_{- 15}$& $   35^{+   12}_{-    8}$& $0.63^{+0.08}_{-0.08}$& $76.0^{+ 8.9}_{- 8.8}$& $3.13^{+0.80}_{-0.38}$& $   2528 $& $    -24555$\\
  &     116  & $ -117^{+ 15}_{- 14}$& $   30^{+ 18}_{- 19}$& $  121^{+ 16}_{- 13}$& $  134^{+  8}_{-  9}$& $   31^{+    9}_{-    7}$& $0.62^{+0.08}_{-0.08}$& $75.3^{+ 8.4}_{- 9.5}$& $2.14^{+0.16}_{-0.16}$& $   2568 $& $    -40971$\\
  &     115  & $ -120^{+ 14}_{- 18}$& $   30^{+ 19}_{- 20}$& $  124^{+ 21}_{- 12}$& $  130^{+  8}_{-  8}$& $   28^{+    9}_{-    7}$& $0.65^{+0.07}_{-0.09}$& $75.7^{+ 8.0}_{- 9.6}$& $1.79^{+0.15}_{-0.12}$& $   2555 $& $    -55354$\\
DraI      &      79  & $    6^{+  6}_{-  8}$& $   48^{+ 14}_{- 11}$& $  159^{+ 11}_{- 10}$& $  106^{+ 13}_{- 10}$& $   42^{+    6}_{-    7}$& $0.43^{+0.06}_{-0.03}$& $73.4^{+ 4.6}_{- 3.4}$& $2.26^{+0.37}_{-0.31}$& $   3122 $& $    -29515$\\
  &      80  & $    1^{+  5}_{-  8}$& $   48^{+ 12}_{- 16}$& $  159^{+ 12}_{- 11}$& $   97^{+  8}_{-  5}$& $   39^{+    6}_{-    5}$& $0.43^{+0.05}_{-0.05}$& $73.0^{+ 4.0}_{- 5.4}$& $1.70^{+0.16}_{-0.10}$& $   3110 $& $    -48412$\\
  &      79  & $    4^{+  7}_{-  8}$& $   48^{+ 12}_{- 10}$& $  159^{+ 14}_{- 10}$& $   90^{+  6}_{-  5}$& $   34^{+    5}_{-    4}$& $0.45^{+0.04}_{-0.04}$& $73.1^{+ 4.0}_{- 3.4}$& $1.38^{+0.11}_{-0.08}$& $   3118 $& $    -66704$\\
DraII     &      22  & $   69^{+  7}_{-  8}$& $   72^{+  6}_{-  8}$& $  348^{+  7}_{-  9}$& $  131^{+ 25}_{- 27}$& $   19^{+    1}_{-    1}$& $0.75^{+0.02}_{-0.05}$& $79.6^{+ 0.9}_{- 1.0}$& $2.37^{+0.58}_{-0.60}$& $   1272 $& $    -28537$\\
  &      22  & $   65^{+  7}_{-  6}$& $   69^{+  7}_{-  7}$& $  347^{+ 10}_{-  8}$& $  114^{+ 16}_{- 10}$& $   19^{+    1}_{-    1}$& $0.71^{+0.03}_{-0.02}$& $79.9^{+ 1.0}_{- 1.0}$& $1.73^{+0.23}_{-0.14}$& $   1223 $& $    -47113$\\
  &      22  & $   68^{+  7}_{-  8}$& $   71^{+  6}_{-  7}$& $  348^{+  7}_{- 10}$& $   79^{+  6}_{-  8}$& $   19^{+    1}_{-    1}$& $0.61^{+0.03}_{-0.03}$& $79.6^{+ 0.9}_{- 1.0}$& $1.10^{+0.07}_{-0.10}$& $   1263 $& $    -75272$\\
EriII     &     364  & $  -37^{+   80}_{-   80}$& $  770^{+   86}_{-   89}$& $  772^{+   90}_{-   80}$& --- & --- & --- & --- & ---  & $ 175224 $& $    283935$\\
  &     366  & $  -33^{+   96}_{-   85}$& $  768^{+   78}_{-   98}$& $  770^{+   84}_{-   89}$& --- & --- & --- & --- & ---  & $ 174746 $& $    279188$\\
  &     365  & $  -35^{+   64}_{-   77}$& $  769^{+   83}_{-  112}$& $  771^{+   86}_{-  102}$& --- & --- & --- & --- & ---  & $ 175075 $& $    275003$\\
\hline
 \end{tabular}
 \end{scriptsize}
  \end{center}
  \end{table*}

\begin{table*}
 \begin{center}
\centerline{\small {\bf Table~4.} Continued from previous page.  }
\begin{scriptsize}
  \begin{tabular}{|l|r|r|r|r|r|r|r|r|r|r|r|}\hline
 Name &$d_{GC}$&$\Pi$ &$\Theta$ &$V_{tot}$ &apo& peri &ecc&incl. $\theta$&$T_r$&$L_Z$&$E$   \\
    &[kpc]&[km/s]&[km/s]&[km/s]& [kpc]& [kpc]&  &[deg]&[Gyr]&[kpc km/s]&[(km/s)$^2$]\\\hline
FnxI      &     141  & $   57^{+ 28}_{- 23}$& $  -89^{+ 30}_{- 28}$& $  128^{+ 32}_{- 24}$& $  160^{+ 83}_{- 36}$& $   88^{+   37}_{-   35}$& $0.29^{+0.24}_{-0.09}$& $108.9^{+ 4.6}_{- 3.1}$& $4.48^{+2.98}_{-1.53}$& $  -5561 $& $    -20238$\\
  &     142  & $   61^{+ 31}_{- 26}$& $  -93^{+ 22}_{- 34}$& $  132^{+ 36}_{- 16}$& $  147^{+ 23}_{- 11}$& $   72^{+   39}_{-   18}$& $0.34^{+0.13}_{-0.18}$& $108.9^{+ 3.4}_{- 3.1}$& $2.79^{+0.75}_{-0.31}$& $  -5768 $& $    -33514$\\
  &     140  & $   58^{+ 30}_{- 26}$& $  -91^{+ 31}_{- 22}$& $  130^{+ 26}_{- 23}$& $  145^{+  9}_{- 10}$& $   57^{+   23}_{-   20}$& $0.43^{+0.15}_{-0.14}$& $108.9^{+ 4.9}_{- 2.9}$& $2.26^{+0.28}_{-0.24}$& $  -5652 $& $    -46767$\\
GruI      &     116  & $   57^{+   26}_{-   28}$& $ -128^{+   31}_{-   24}$& $  300^{+   21}_{-   16}$& --- & --- & --- & --- & ---  & $  -7043 $& $     12422$\\
  &     116  & $   59^{+   20}_{-   30}$& $ -123^{+   20}_{-   22}$& $  299^{+   13}_{-   18}$& --- & --- & --- & --- & ---  & $  -6800 $& $     -3914$\\
  &     116  & $   58^{+ 26}_{- 30}$& $ -126^{+ 35}_{- 27}$& $  299^{+ 21}_{- 16}$& $  400^{+188}_{-186}$& $   69^{+   12}_{-   43}$& $0.71^{+0.18}_{-0.08}$& $106.0^{+ 3.7}_{- 2.7}$& $7.27^{+5.07}_{-4.15}$& $  -6940 $& $    -18450$\\
HerI      &     128  & $  110^{+ 19}_{- 19}$& $   51^{+ 32}_{- 24}$& $  159^{+ 22}_{- 13}$& $  265^{+118}_{- 62}$& $   16^{+   17}_{-    7}$& $0.88^{+0.06}_{-0.09}$& $40.4^{+ 5.3}_{-17.9}$& $5.54^{+3.82}_{-1.83}$& $   5145 $& $    -17683$\\
  &     130  & $  107^{+ 22}_{- 21}$& $   55^{+ 27}_{- 24}$& $  159^{+ 21}_{- 12}$& $  187^{+ 27}_{- 16}$& $   18^{+   15}_{-    9}$& $0.83^{+0.07}_{-0.13}$& $40.6^{+ 3.8}_{-11.1}$& $2.75^{+0.48}_{-0.21}$& $   5574 $& $    -32521$\\
  &     130  & $  109^{+ 20}_{- 21}$& $   53^{+ 26}_{- 27}$& $  159^{+ 19}_{- 13}$& $  172^{+ 27}_{- 29}$& $   15^{+   12}_{-    8}$& $0.84^{+0.08}_{-0.11}$& $40.5^{+ 4.3}_{-15.6}$& $2.19^{+0.30}_{-0.23}$& $   5312 $& $    -46313$\\
HorI      &      80  & $  131^{+ 30}_{- 28}$& $    5^{+ 17}_{- 25}$& $  191^{+ 26}_{- 21}$& $  127^{+118}_{- 52}$& $   74^{+    9}_{-   16}$& $0.26^{+0.23}_{-0.07}$& $89.1^{+ 3.2}_{- 4.5}$& $3.38^{+3.27}_{-1.59}$& $    223 $& $    -23692$\\
  &      79  & $  136^{+ 25}_{- 28}$& $    4^{+ 24}_{- 22}$& $  195^{+ 22}_{- 18}$& $   98^{+ 93}_{- 67}$& $   73^{+    8}_{-   18}$& $0.15^{+0.20}_{-0.07}$& $89.3^{+ 4.4}_{- 4.0}$& $2.14^{+2.64}_{-2.14}$& $    181 $& $    -41894$\\
  &      79  & $  133^{+ 24}_{- 30}$& $    4^{+ 18}_{- 27}$& $  193^{+ 22}_{- 20}$& $   86^{+ 21}_{- 11}$& $   63^{+    9}_{-   13}$& $0.15^{+0.14}_{-0.06}$& $89.3^{+ 3.4}_{- 4.9}$& $1.63^{+0.24}_{-0.22}$& $    194 $& $    -60536$\\
HorII     &      77  & $  319^{+  106}_{-   91}$& $   62^{+  145}_{-  113}$& $  383^{+  129}_{-   57}$& --- & --- & --- & --- & ---  & $   2876 $& $     30483$\\
  &      78  & $  324^{+  101}_{-   74}$& $   60^{+  139}_{-   90}$& $  387^{+  116}_{-   51}$& --- & --- & --- & --- & ---  & $   2801 $& $     12872$\\
  &      77  & $  321^{+   99}_{-   95}$& $   61^{+   98}_{-  136}$& $  384^{+  108}_{-   65}$& --- & --- & --- & --- & ---  & $   2819 $& $     -6486$\\
HyaII     &     132  & $   11^{+   33}_{-   19}$& $  160^{+   28}_{-   49}$& $  264^{+   25}_{-   28}$& --- & --- & --- & --- & ---  & $  17561 $& $      5052$\\
  &     133  & $   16^{+ 26}_{- 29}$& $  161^{+ 38}_{- 39}$& $  266^{+ 40}_{- 35}$& $  593^{+  0}_{-590}$& $  105^{+   14}_{-   16}$& $0.70^{+0.28}_{-0.17}$& $54.5^{+ 5.0}_{- 6.0}$& $15.53^{+0.00}_{-15.78}$& $  17744 $& $     -9216$\\
  &     133  & $   13^{+ 29}_{- 26}$& $  160^{+ 29}_{- 37}$& $  265^{+ 32}_{- 24}$& $  283^{+238}_{-114}$& $  101^{+   13}_{-   16}$& $0.47^{+0.21}_{-0.07}$& $54.8^{+ 4.4}_{- 6.6}$& $5.07^{+6.94}_{-3.10}$& $  17626 $& $    -23038$\\
HyiI      &      26  & $  163^{+ 28}_{- 27}$& $  166^{+ 18}_{- 13}$& $  370^{+ 22}_{- 21}$& $  362^{+113}_{-312}$& $   26^{+    3}_{-   22}$& $0.87^{+0.12}_{-0.09}$& $69.6^{+ 2.6}_{- 2.0}$& $8.49^{+3.90}_{-8.04}$& $   3301 $& $    -13882$\\
  &      26  & $  168^{+ 28}_{- 29}$& $  166^{+ 16}_{- 11}$& $  372^{+ 25}_{- 20}$& $  188^{+106}_{- 55}$& $   26^{+    1}_{-    1}$& $0.76^{+0.07}_{-0.06}$& $69.73^{+ 2.4}_{- 1.8}$& $2.83^{+2.33}_{-1.09}$& $   3306 $& $    -31903$\\
  &      26  & $  165^{+ 28}_{- 32}$& $  165^{+ 16}_{- 12}$& $  371^{+ 23}_{- 15}$& $  116^{+ 37}_{- 19}$& $   26^{+    1}_{-    2}$& $0.64^{+0.07}_{-0.04}$& $69.8^{+ 2.2}_{- 2.0}$& $1.60^{+0.46}_{-0.23}$& $   3285 $& $    -59919$\\
LeoI      &     272  & $  131^{+   47}_{-   38}$& $   77^{+   42}_{-   46}$& $  186^{+   47}_{-   25}$& --- & --- & --- & --- & ---  & $  13957 $& $      -323$\\
  &     274  & $  134^{+   39}_{-   47}$& $   73^{+   48}_{-   22}$& $  187^{+   45}_{-   26}$& --- & --- & --- & --- & ---  & $  13301 $& $     -5430$\\
  &     275  & $  132^{+ 41}_{- 32}$& $   75^{+ 43}_{- 40}$& $  186^{+ 44}_{- 20}$& $  620^{+114}_{-533}$& $   53^{+   44}_{-   18}$& $0.84^{+0.15}_{-0.06}$& $51.0^{+ 6.5}_{-19.3}$& $12.95^{+3.52}_{-10.24}$& $  13699 $& $    -12571$\\
LeoII     &     229  & $  -62^{+ 36}_{- 29}$& $   29^{+ 31}_{- 39}$& $   84^{+ 32}_{- 17}$& $  232^{+ 61}_{- 21}$& $   87^{+   77}_{-   42}$& $0.46^{+0.22}_{-0.24}$& $81.5^{+ 8.4}_{-11.3}$& $6.28^{+3.61}_{-1.25}$& $   2735 $& $    -16671$\\
  &     229  & $  -59^{+ 42}_{- 23}$& $   25^{+ 39}_{- 36}$& $   80^{+ 31}_{- 20}$& $  229^{+ 18}_{- 11}$& $   58^{+   39}_{-   28}$& $0.60^{+0.17}_{-0.19}$& $82.3^{+10.5}_{-10.7}$& $3.88^{+0.68}_{-0.32}$& $   2362 $& $    -24281$\\
  &     228  & $  -61^{+ 38}_{- 42}$& $   28^{+ 31}_{- 39}$& $   82^{+ 41}_{- 16}$& $  228^{+ 56}_{- 39}$& $   50^{+   43}_{-   20}$& $0.64^{+0.12}_{-0.21}$& $81.8^{+ 7.8}_{-12.3}$& $3.30^{+1.55}_{-0.85}$& $   2596 $& $    -32572$\\
LeoIV     &     153  & $  266^{+   42}_{-   46}$& $  173^{+   43}_{-   36}$& $  354^{+   43}_{-   35}$& --- & --- & --- & --- & ---  & $  14589 $& $     35941$\\
  &     155  & $  270^{+   34}_{-   56}$& $  172^{+   45}_{-   55}$& $  357^{+   39}_{-   51}$& --- & --- & --- & --- & ---  & $  14498 $& $     24405$\\
  &     155  & $  268^{+   30}_{-   58}$& $  172^{+   38}_{-   46}$& $  355^{+   31}_{-   50}$& --- & --- & --- & --- & ---  & $  14532 $& $     11689$\\
LeoV      &     174  & $  213^{+   50}_{-   45}$& $ -225^{+   46}_{-   52}$& $  322^{+   54}_{-   39}$& --- & --- & --- & --- & ---  & $ -20697 $& $     27287$\\
  &     175  & $  218^{+   39}_{-   56}$& $ -226^{+   49}_{-   59}$& $  326^{+   50}_{-   44}$& --- & --- & --- & --- & ---  & $ -20812 $& $     17495$\\
  &     173  & $  215^{+   43}_{-   55}$& $ -225^{+   49}_{-   54}$& $  323^{+   51}_{-   45}$& --- & --- & --- & --- & ---  & $ -20768 $& $      5753$\\
LMC       &      50  & $  211^{+ 18}_{- 18}$& $   44^{+ 12}_{-  9}$& $  307^{+ 21}_{- 17}$& $  578^{+  0}_{-587}$& $   49^{+    5}_{-    6}$& $0.85^{+0.15}_{-0.11}$& $82.9^{+ 2.0}_{- 1.5}$& $16.20^{+0.00}_{-16.92}$& $   1846 $& $     -9574$\\
  &      50  & $  216^{+ 21}_{- 16}$& $   44^{+  8}_{- 11}$& $  310^{+ 25}_{- 21}$& $  196^{+ 95}_{- 59}$& $   48^{+    2}_{-    3}$& $0.60^{+0.11}_{-0.07}$& $83.1^{+ 1.5}_{- 1.9}$& $3.18^{+2.19}_{-1.17}$& $   1831 $& $    -28870$\\
  &      50  & $  213^{+ 20}_{- 21}$& $   44^{+  9}_{- 10}$& $  308^{+ 25}_{- 23}$& $  130^{+ 37}_{- 25}$& $   48^{+    2}_{-    3}$& $0.46^{+0.09}_{-0.07}$& $83.1^{+ 1.8}_{- 1.7}$& $1.99^{+0.46}_{-0.32}$& $   1819 $& $    -51851$\\
PheII     &      77  & $  126^{+ 51}_{- 54}$& $  173^{+ 64}_{- 37}$& $  240^{+ 69}_{- 38}$& $  316^{+ 86}_{-344}$& $   75^{+    0}_{-   76}$& $0.62^{+0.42}_{-0.16}$& $71.6^{+ 4.2}_{- 3.3}$& $8.36^{+3.18}_{-9.53}$& $   5749 $& $    -14111$\\
  &      76  & $  129^{+ 47}_{- 54}$& $  176^{+ 60}_{- 39}$& $  244^{+ 63}_{- 42}$& $  155^{+134}_{- 91}$& $   75^{+   10}_{-   27}$& $0.35^{+0.33}_{-0.11}$& $71.6^{+ 3.8}_{- 2.5}$& $2.92^{+3.02}_{-1.94}$& $   5862 $& $    -32074$\\
  &      77  & $  128^{+ 57}_{- 56}$& $  174^{+ 56}_{- 46}$& $  241^{+ 66}_{- 40}$& $  113^{+151}_{- 58}$& $   73^{+    7}_{-   21}$& $0.21^{+0.30}_{-0.03}$& $71.6^{+ 3.5}_{- 3.9}$& $2.05^{+2.90}_{-1.43}$& $   5771 $& $    -51261$\\
PhxI      &     423  & $  -95^{+   69}_{-   73}$& $   53^{+   71}_{-   89}$& $  141^{+   78}_{-   18}$& --- & --- & --- & --- & ---  & $   7744 $& $     -2847$\\
  &     423  & $  -90^{+   79}_{-   74}$& $   53^{+   70}_{-   74}$& $  137^{+   82}_{-   26}$& --- & --- & --- & --- & ---  & $   7691 $& $     -5512$\\
  &     421  & $  -93^{+   75}_{-   81}$& $   53^{+   66}_{-   85}$& $  139^{+   87}_{-   36}$& --- & --- & --- & --- & ---  & $   7733 $& $     -9814$\\
PisII     &     182  & $ -174^{+   33}_{-   57}$& $  333^{+   49}_{-   62}$& $  381^{+   58}_{-   46}$& --- & --- & --- & --- & ---  & $  41039 $& $     48805$\\
  &     183  & $ -179^{+   47}_{-   46}$& $  334^{+   68}_{-   58}$& $  384^{+   66}_{-   45}$& --- & --- & --- & --- & ---  & $  41128 $& $     39436$\\
  &     182  & $ -176^{+   38}_{-   58}$& $  334^{+   53}_{-   49}$& $  382^{+   63}_{-   38}$& --- & --- & --- & --- & ---  & $  41051 $& $     28240$\\
RetII     &      32  & $   70^{+ 19}_{- 19}$& $  -42^{+ 11}_{- 12}$& $  226^{+ 21}_{- 18}$& $   50^{+ 13}_{-  8}$& $   22^{+    3}_{-    3}$& $0.39^{+0.10}_{-0.05}$& $98.3^{+ 1.9}_{- 1.8}$& $0.91^{+0.24}_{-0.16}$& $   -914 $& $    -48332$\\
  &      32  & $   75^{+ 16}_{- 18}$& $  -44^{+ 11}_{-  9}$& $  227^{+ 16}_{- 16}$& $   52^{+ 10}_{-  7}$& $   24^{+    2}_{-    3}$& $0.38^{+0.07}_{-0.06}$& $98.6^{+ 1.9}_{- 1.4}$& $0.93^{+0.14}_{-0.10}$& $   -963 $& $    -67592$\\
  &      32  & $   72^{+ 19}_{- 18}$& $  -43^{+ 12}_{- 10}$& $  226^{+ 23}_{- 19}$& $   45^{+  7}_{-  5}$& $   22^{+    3}_{-    3}$& $0.34^{+0.08}_{-0.05}$& $98.5^{+ 1.9}_{- 1.5}$& $0.72^{+0.10}_{-0.07}$& $   -941 $& $    -93834$\\
\hline
 \end{tabular}
 \end{scriptsize}
  \end{center}
  \end{table*}

\begin{table*}
 \begin{center}
\centerline{\small {\bf Table~4.} Continued from previous page.  }
\begin{scriptsize}
 \begin{tabular}{|l|r|r|r|r|r|r|r|r|r|r|r|}\hline
 Name &$d_{GC}$&$\Pi$ &$\Theta$ &$V_{tot}$ &apo& peri &ecc&incl. $\theta$&$T_r$&$L_z$&$E$   \\
    &[kpc]&[km/s]&[km/s]&[km/s]& [kpc]& [kpc]&  &[deg]&[Gyr]&[kpc km/s]&[(km/s)$^2$]\\\hline
RetIII    &      94  & $ -154^{+  181}_{-  166}$& $ -201^{+  298}_{-  167}$& $  355^{+  219}_{-   83}$& --- & --- & --- & --- & ---  & $ -12922 $& $     25054$\\
  &      92  & $ -149^{+  156}_{-  169}$& $ -202^{+  290}_{-  267}$& $  354^{+  310}_{-   40}$& --- & --- & --- & --- & ---  & $ -12956 $& $      6557$\\
  &      91  & $ -152^{+153}_{-174}$& $ -201^{+281}_{-256}$& $  355^{+319}_{- 48}$& $  753^{+  0}_{-819}$& $   89^{+    6}_{-   63}$& $0.79^{+0.32}_{-0.15}$& $113.8^{+31.7}_{-13.7}$& $18.20^{+0.00}_{-19.55}$& $ -12923 $& $     -9954$\\
SgrI      &      18  & $  225^{+  6}_{-  8}$& $   44^{+ 22}_{- 15}$& $  310^{+ 12}_{- 10}$& $   50^{+ 12}_{- 11}$& $   14^{+    1}_{-    1}$& $0.55^{+0.05}_{-0.05}$& $81.4^{+ 4.1}_{- 2.9}$& $0.79^{+0.22}_{-0.18}$& $    746 $& $    -51706$\\
  &      18  & $  224^{+  6}_{-  7}$& $   49^{+ 17}_{- 21}$& $  310^{+ 11}_{- 12}$& $   57^{+ 12}_{- 11}$& $   15^{+    1}_{-    1}$& $0.59^{+0.06}_{-0.05}$& $80.5^{+ 3.4}_{- 4.1}$& $0.89^{+0.16}_{-0.16}$& $    829 $& $    -68513$\\
  &      18  & $  225^{+  6}_{-  8}$& $   46^{+ 21}_{- 18}$& $  310^{+  9}_{- 10}$& $   44^{+  7}_{-  8}$& $   14^{+    1}_{-    1}$& $0.51^{+0.06}_{-0.06}$& $81.0^{+ 4.2}_{- 3.4}$& $0.63^{+0.08}_{-0.10}$& $    777 $& $    -98704$\\
SgrII     &      60  & $  -45^{+ 22}_{- 26}$& $   93^{+ 41}_{- 34}$& $  190^{+ 36}_{- 29}$& $   96^{+ 65}_{- 30}$& $   33^{+    9}_{-    8}$& $0.49^{+0.13}_{-0.06}$& $57.5^{+12.9}_{-11.0}$& $1.89^{+1.65}_{-0.79}$& $   4989 $& $    -32506$\\
  &      60  & $  -46^{+ 22}_{- 21}$& $   97^{+ 36}_{- 41}$& $  193^{+ 36}_{- 28}$& $   87^{+ 29}_{- 14}$& $   34^{+    8}_{-    7}$& $0.45^{+0.10}_{-0.06}$& $56.2^{+10.6}_{-13.5}$& $1.52^{+0.46}_{-0.26}$& $   5237 $& $    -52009$\\
  &      60  & $  -45^{+ 27}_{- 24}$& $   94^{+ 40}_{- 44}$& $  191^{+ 37}_{- 31}$& $   77^{+ 17}_{-  9}$& $   30^{+    8}_{-    6}$& $0.45^{+0.09}_{-0.07}$& $57.0^{+11.4}_{-14.6}$& $1.18^{+0.26}_{-0.16}$& $   5075 $& $    -73168$\\
SclI      &      84  & $ -150^{+ 16}_{- 17}$& $  106^{+ 18}_{- 14}$& $  208^{+ 15}_{- 11}$& $  204^{+103}_{- 59}$& $   71^{+    6}_{-    7}$& $0.49^{+0.09}_{-0.07}$& $86.2^{+ 0.6}_{- 0.5}$& $5.16^{+3.13}_{-1.77}$& $   1097 $& $    -18618$\\
  &      85  & $ -146^{+ 17}_{- 16}$& $  103^{+ 16}_{- 15}$& $  203^{+ 14}_{- 12}$& $  125^{+ 18}_{- 12}$& $   66^{+    7}_{-    7}$& $0.31^{+0.03}_{-0.00}$& $86.2^{+ 0.5}_{- 0.5}$& $2.42^{+0.31}_{-0.25}$& $   1065 $& $    -37934$\\
  &      85  & $ -147^{+ 14}_{- 15}$& $  106^{+ 13}_{- 17}$& $  206^{+ 13}_{- 13}$& $  108^{+ 10}_{-  9}$& $   61^{+    6}_{-    7}$& $0.28^{+0.03}_{-0.02}$& $86.1^{+ 0.4}_{- 0.5}$& $1.86^{+0.17}_{-0.17}$& $   1103 $& $    -55036$\\
Seg1      &      28  & $  184^{+ 12}_{- 13}$& $ -133^{+ 25}_{- 24}$& $  230^{+ 23}_{- 22}$& $   46^{+ 12}_{-  8}$& $   18^{+    3}_{-    2}$& $0.43^{+0.07}_{-0.04}$& $121.6^{+ 4.4}_{- 3.2}$& $0.78^{+0.22}_{-0.15}$& $  -2900 $& $    -52276$\\
  &      28  & $  186^{+ 12}_{- 14}$& $ -138^{+ 26}_{- 23}$& $  235^{+ 23}_{- 25}$& $   50^{+ 12}_{-  9}$& $   19^{+    2}_{-    3}$& $0.44^{+0.07}_{-0.04}$& $122.1^{+ 4.1}_{- 3.1}$& $0.84^{+0.18}_{-0.14}$& $  -2997 $& $    -70512$\\
  &      28  & $  185^{+ 14}_{- 11}$& $ -135^{+ 21}_{- 25}$& $  232^{+ 26}_{- 19}$& $   42^{+  8}_{-  4}$& $   18^{+    3}_{-    2}$& $0.40^{+0.06}_{-0.03}$& $122.0^{+ 3.5}_{- 3.6}$& $0.65^{+0.11}_{-0.07}$& $  -2958 $& $    -97922$\\
Seg2      &      42  & $  184^{+ 16}_{- 15}$& $   -1^{+ 17}_{- 16}$& $  233^{+ 14}_{- 13}$& $   79^{+ 20}_{- 20}$& $   35^{+    2}_{-    3}$& $0.38^{+0.14}_{-0.10}$& $90.37^{+ 3.9}_{- 3.5}$& $1.61^{+0.37}_{-0.37}$& $    -46 $& $    -35557$\\
  &      42  & $  182^{+ 15}_{- 13}$& $   -6^{+ 17}_{- 14}$& $  231^{+ 14}_{- 10}$& $   72^{+ 17}_{- 18}$& $   36^{+    2}_{-    2}$& $0.34^{+0.17}_{-0.11}$& $91.3^{+ 3.9}_{- 3.0}$& $1.33^{+0.25}_{-0.26}$& $   -204 $& $    -56221$\\
  &      42  & $  184^{+ 17}_{- 13}$& $   -3^{+ 15}_{- 14}$& $  232^{+ 15}_{- 11}$& $   59^{+ 56}_{- 43}$& $   34^{+    2}_{-    3}$& $0.28^{+0.25}_{-0.12}$& $90.6^{+ 3.4}_{- 3.3}$& $1.01^{+0.71}_{-0.60}$& $   -100 $& $    -79686$\\
SMC       &      62  & $  170^{+ 18}_{- 26}$& $   68^{+ 11}_{- 16}$& $  244^{+ 19}_{- 26}$& $  191^{+106}_{- 98}$& $   61^{+    5}_{-    8}$& $0.51^{+0.17}_{-0.20}$& $79.2^{+ 1.8}_{- 2.3}$& $4.60^{+3.04}_{-2.52}$& $   2796 $& $    -19852$\\
  &      61  & $  174^{+ 23}_{- 25}$& $   70^{+ 18}_{- 14}$& $  248^{+ 21}_{- 24}$& $  123^{+ 31}_{- 30}$& $   61^{+    2}_{-    4}$& $0.34^{+0.10}_{-0.12}$& $79.1^{+ 2.9}_{- 1.8}$& $2.34^{+0.42}_{-0.43}$& $   2869 $& $    -38890$\\
  &      61  & $  172^{+ 24}_{- 24}$& $   69^{+ 14}_{- 15}$& $  245^{+ 24}_{- 23}$& $   89^{+ 26}_{- 18}$& $   61^{+    3}_{-    5}$& $0.18^{+0.12}_{-0.07}$& $79.3^{+ 2.0}_{- 2.2}$& $1.65^{+0.30}_{-0.23}$& $   2806 $& $    -60004$\\
SxtI      &      89  & $   46^{+   15}_{-   21}$& $  242^{+   13}_{-   18}$& $  256^{+   15}_{-   19}$& --- & --- & --- & --- & ---  & $  16224 $& $     -6275$\\
  &      89  & $   50^{+ 13}_{- 21}$& $  239^{+ 16}_{- 12}$& $  254^{+ 17}_{- 12}$& $  207^{+ 59}_{- 31}$& $   80^{+    5}_{-    4}$& $0.44^{+0.09}_{-0.05}$& $41.3^{+ 2.6}_{- 1.9}$& $3.71^{+1.38}_{-0.60}$& $  16033 $& $    -24883$\\
  &      89  & $   48^{+ 17}_{- 16}$& $  241^{+ 14}_{- 16}$& $  255^{+ 15}_{- 16}$& $  154^{+ 24}_{- 21}$& $   78^{+    5}_{-    5}$& $0.33^{+0.06}_{-0.06}$& $41.3^{+ 2.2}_{- 2.6}$& $2.58^{+0.30}_{-0.29}$& $  16160 $& $    -41783$\\
TriII     &      36  & $ -171^{+  8}_{-  7}$& $  124^{+  9}_{-  9}$& $  331^{+  9}_{-  7}$& $  376^{+146}_{-105}$& $   20^{+    4}_{-   11}$& $0.90^{+0.05}_{-0.03}$& $58.1^{+ 1.9}_{- 2.3}$& $8.79^{+4.57}_{-2.89}$& $   4246 $& $    -13590$\\
  &      36  & $ -174^{+ 10}_{-  8}$& $  120^{+  9}_{-  9}$& $  331^{+  8}_{- 10}$& $  179^{+ 19}_{- 21}$& $   20^{+    1}_{-    2}$& $0.80^{+0.02}_{-0.02}$& $58.8^{+ 2.2}_{- 1.9}$& $2.66^{+0.28}_{-0.32}$& $   4104 $& $    -33709$\\
  &      37  & $ -172^{+  9}_{-  9}$& $  122^{+ 10}_{-  8}$& $  331^{+  9}_{-  8}$& $  124^{+ 12}_{-  9}$& $   20^{+    2}_{-    1}$& $0.73^{+0.02}_{-0.02}$& $58.4^{+ 2.2}_{- 2.0}$& $1.64^{+0.16}_{-0.11}$& $   4192 $& $    -58542$\\
TucII     &      53  & $ -150^{+ 28}_{- 22}$& $  201^{+ 26}_{- 22}$& $  282^{+ 16}_{- 13}$& $  335^{+189}_{-162}$& $   33^{+    6}_{-   20}$& $0.82^{+0.11}_{-0.06}$& $60.4^{+ 2.5}_{- 2.3}$& $7.85^{+6.22}_{-4.59}$& $   5530 $& $    -14538$\\
  &      53  & $ -147^{+ 23}_{- 22}$& $  204^{+ 25}_{- 18}$& $  283^{+ 17}_{- 10}$& $  167^{+ 37}_{- 19}$& $   34^{+    4}_{-    3}$& $0.66^{+0.04}_{-0.02}$& $60.2^{+ 2.6}_{- 2.5}$& $2.63^{+0.61}_{-0.34}$& $   5631 $& $    -34531$\\
  &      54  & $ -149^{+ 23}_{- 22}$& $  203^{+ 26}_{- 19}$& $  282^{+ 17}_{- 10}$& $  124^{+ 21}_{- 10}$& $   32^{+    5}_{-    3}$& $0.59^{+0.04}_{-0.02}$& $60.4^{+ 2.8}_{- 2.2}$& $1.75^{+0.29}_{-0.14}$& $   5564 $& $    -56511$\\
TucIII    &      23  & $ -130^{+ 10}_{- 13}$& $  -44^{+ 11}_{- 10}$& $  236^{+ 10}_{-  8}$& $   46^{+  6}_{-  5}$& $    3^{+    1}_{-    1}$& $0.89^{+0.03}_{-0.02}$& $108.5^{+ 5.0}_{- 4.6}$& $0.59^{+0.08}_{-0.06}$& $   -425 $& $    -59206$\\
  &      23  & $ -125^{+ 13}_{-  8}$& $  -43^{+  9}_{- 11}$& $  233^{+  8}_{- 10}$& $   47^{+ 88}_{- 72}$& $    3^{+    1}_{-    1}$& $0.90^{+0.02}_{-0.02}$& $109.4^{+ 3.9}_{- 5.0}$& $0.63^{+0.06}_{-0.06}$& $   -419 $& $    -78158$\\
  &      23  & $ -128^{+ 14}_{- 11}$& $  -42^{+ 12}_{-  8}$& $  235^{+ 10}_{- 10}$& $   42^{+ 10}_{-  7}$& $    3^{+    1}_{-    1}$& $0.89^{+0.03}_{-0.02}$& $108.5^{+ 5.0}_{- 4.7}$& $0.51^{+0.05}_{-0.04}$& $   -409 $& $   -106063$\\
UMaI      &     103  & $  145^{+   25}_{-   24}$& $ -199^{+   28}_{-   30}$& $  264^{+   26}_{-   19}$& --- & --- & --- & --- & ---  & $ -12791 $& $      -719$\\
  &     103  & $  143^{+ 26}_{- 21}$& $ -203^{+ 37}_{- 23}$& $  267^{+ 21}_{- 26}$& $  292^{+134}_{-150}$& $  102^{+   13}_{-   17}$& $0.48^{+0.21}_{-0.17}$& $118.8^{+ 4.2}_{- 2.3}$& $6.07^{+4.06}_{-3.71}$& $ -13089 $& $    -17091$\\
  &     102  & $  145^{+ 18}_{- 21}$& $ -200^{+ 26}_{- 34}$& $  265^{+ 26}_{- 18}$& $  184^{+ 74}_{- 38}$& $  102^{+    4}_{-    5}$& $0.29^{+0.13}_{-0.08}$& $118.6^{+ 2.9}_{- 2.8}$& $3.18^{+1.52}_{-0.74}$& $ -12921 $& $    -33461$\\
UMaII     &      41  & $ -187^{+  7}_{- 11}$& $  -30^{+ 18}_{- 24}$& $  284^{+ 20}_{- 16}$& $  163^{+127}_{- 57}$& $   40^{+    5}_{-    7}$& $0.61^{+0.13}_{-0.10}$& $95.3^{+ 3.1}_{- 4.0}$& $3.46^{+3.53}_{-1.52}$& $  -1065 $& $    -23240$\\
  &      41  & $ -188^{+  8}_{- 11}$& $  -35^{+ 19}_{- 20}$& $  286^{+ 17}_{- 12}$& $  119^{+ 29}_{- 17}$& $   40^{+    1}_{-    2}$& $0.50^{+0.07}_{-0.05}$& $96.1^{+ 3.2}_{- 3.4}$& $2.02^{+0.34}_{-0.23}$& $  -1225 $& $    -43016$\\
  &      41  & $ -187^{+ 12}_{-  8}$& $  -32^{+ 23}_{- 18}$& $  285^{+ 17}_{- 18}$& $   85^{+ 21}_{- 15}$& $   40^{+    2}_{-    2}$& $0.36^{+0.07}_{-0.06}$& $95.7^{+ 4.1}_{- 3.1}$& $1.37^{+0.23}_{-0.21}$& $  -1137 $& $    -67198$\\
UMiI      &      78  & $   54^{+  4}_{-  4}$& $   23^{+  6}_{-  5}$& $  160^{+  4}_{-  4}$& $   98^{+  8}_{-  6}$& $   49^{+    4}_{-    4}$& $0.33^{+0.04}_{-0.03}$& $83.4^{+ 1.7}_{- 1.3}$& $2.24^{+0.21}_{-0.16}$& $   1316 $& $    -29689$\\
  &      79  & $   50^{+  4}_{-  4}$& $   21^{+  8}_{-  5}$& $  159^{+  4}_{-  3}$& $   90^{+  6}_{-  4}$& $   44^{+    4}_{-    4}$& $0.34^{+0.04}_{-0.03}$& $83.8^{+ 2.1}_{- 1.5}$& $1.69^{+0.11}_{-0.08}$& $   1213 $& $    -48906$\\
  &      78  & $   52^{+  4}_{-  4}$& $   23^{+  6}_{-  6}$& $  160^{+  3}_{-  3}$& $   86^{+  4}_{-  4}$& $   39^{+    3}_{-    3}$& $0.38^{+0.03}_{-0.03}$& $83.5^{+ 1.8}_{- 1.6}$& $1.37^{+0.06}_{-0.07}$& $   1282 $& $    -67144$\\
Wil1      &      43  & $  -63^{+ 12}_{-  9}$& $   53^{+ 19}_{- 15}$& $  121^{+ 11}_{-  9}$& $   44^{+  3}_{-  1}$& $   16^{+    3}_{-    2}$& $0.46^{+0.05}_{-0.06}$& $72.7^{+ 5.5}_{- 4.8}$& $0.72^{+0.07}_{-0.03}$& $   1523 $& $    -54328$\\
  &      43  & $  -65^{+  9}_{- 13}$& $   48^{+ 14}_{- 21}$& $  120^{+ 12}_{-  7}$& $   44^{+  2}_{-  2}$& $   17^{+    3}_{-    2}$& $0.43^{+0.04}_{-0.07}$& $74.1^{+ 3.9}_{- 6.8}$& $0.74^{+0.06}_{-0.04}$& $   1387 $& $    -74801$\\
  &      43  & $  -64^{+  8}_{- 11}$& $   51^{+ 20}_{- 19}$& $  121^{+ 13}_{-  5}$& $   44^{+  2}_{-  2}$& $   15^{+    3}_{-    1}$& $0.48^{+0.03}_{-0.06}$& $73.2^{+ 5.6}_{- 6.5}$& $0.64^{+0.05}_{-0.02}$& $   1473 $& $    -98215$\\\hline
 \end{tabular}
 \end{scriptsize}
  \end{center}
  \end{table*}

\begin{figure*}
{\begin{center}
   \includegraphics[width=0.9\textwidth,angle=0]{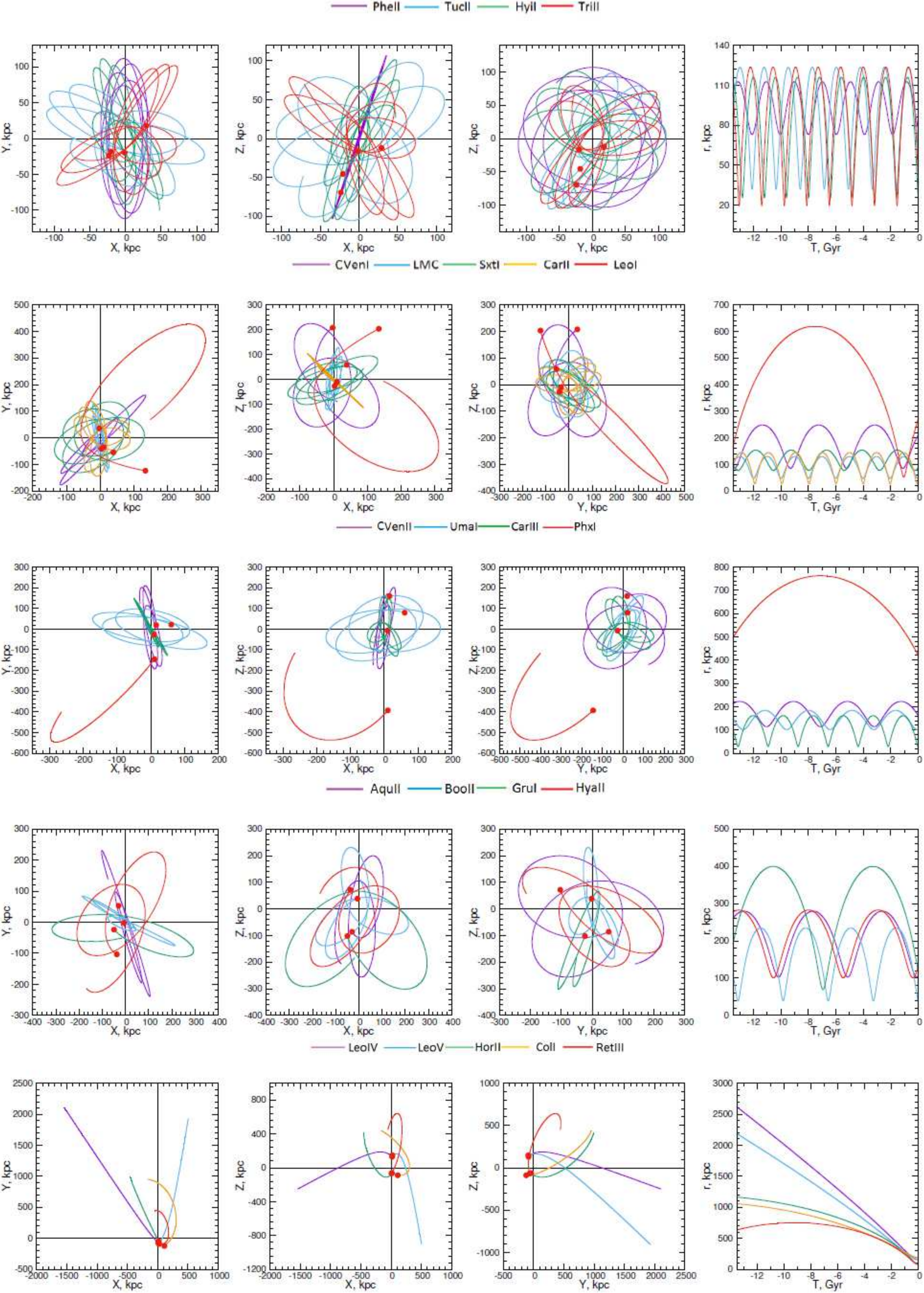}
  \centerline{{\bf Figure~7.} Continued from previous page}
\label{f32}
\end{center}}
\end{figure*}

\section*{Conclusions}

\noindent(i)Based on modern literary data on positions, proper motions (from Gaia DR2 Catalog), and line-of-sight velocities, a sample of 47 dwarf satellite galaxies of the Milky Way was formed to analyze their orbital motion using three models of the Galactic potential with different masses.\\

\noindent(ii)To select the most probable models of the Galactic potential, we collected the Galaxy mass estimates from various literary sources obtained by independent methods.
As a result, three models were selected that describe the axisymmetric potential of the Galaxy: 1)the model with a dark halo in the NFW form, modified by Bajkova \& Bobylev
(2016) (NFWBB), 2)the  Allen \& Santill\'an (1991) model, also modified in Bajkova \& Bobylev (2016) (ASBB) and 3)the  Allen \& Santill\'an (1991) model, modified by Irrgang et
al. (2013) (ASI), with low($M_{G_{(R \leq 200 kpc)}}=0.75\times10^{12}M_\odot$), middle ($M_{G_{(R \leq 200 kpc)}}=1.45\times10^{12}M_\odot$) and high ($M_{G_{(R\leq 200~kpc)}}=1.9\times10^{12}M_\odot$) mass of the Galaxy, respectively.\\

\noindent(iii)We present the orbits and the orbit properties of the dwarf galaxies obtained by integration for 13.5 Gyr backward in the three models of the Galactic potential. We give a comparison of the orbital parameters corresponding to different models of the Galactic potential.\\

\noindent(iv)For each model of the Galactic potential we have identified dwarf galaxies that are not connected gravitationally with the Milky Way. For the NFWBB model these are AquII, BooII, ColI, EriII, GruI, HorII, HyaII, LeoIV, LeoV, PisII, RetIII, for the ASBB model, ColI, EriII, HorII, LeoIV, LeoV, PisII, RetIII), and for the ASI model, EriII, LeoIV, LeoV, PisI.

At this stage of research, it is difficult to say which model of the Galaxy is the most realistic. It is required further refinement of the rotation curve of the Galaxy, especially at large distances from the center of the Galaxy in order to refine the model of the Galaxy and its mass. Further refinement of the three-dimensional dynamics of galactic objects also requires further refinement of astrometric measurements.

\section{References}
\begin{itemize}

\item Ablimit I., Zhao G., Flynn C.,
and Bird S.A., 2020, ApJ Lett. 895, Issue 1, id.L12

\item Aihara H., Prieto C.A., An D., et al., 2011, ApJ Suppl. 193, 29

\item Allen C., Martos M., 1986, RMxAA 13, 137

\item Allen C., Santill\'an  A., 1991, RMxAA 22, 255

\item Bajkova A.T., Bobylev V.V., 2016, Ast. Lett. 42, 567 

\item Bajkova A.T., Bobylev V.V., 2017, OAst 26, 72

\item Bajkova A.T., Bobylev V.V., 2017, Ast. Rep. 61, 727

\item Battaglia G., Helmi A., Morrison H., et al., 2005, MNRAS 364, 433

\item Bhattacharjee P., Chaudhury S., Kundu S., and Majumdar S., 2013,
Phys. Rev. D 87, 083525

\item Bhattacharjee P., Chaudhury S., and Kundu S., 2014, ApJ 785, 63

\item Bobylev V.V., Bajkova A.T., 2016, Ast. Lett. 42, 1     

\item Bobylev V.V., Bajkova A.T., and Gromov A.O., 2017, Ast. Lett. 43, 241  

\item Gaia Collaboration, Brown A.G.A., Vallenari A., Prusti T., et al., 2018, A\&A 616, 1 

\item Callingham T.M., Cautun  M., Deason A.J., et al., 2019, MNRAS 484, 5453

\item Deason A.J., Belokurov V., Evans N.W., and An J., 2012a, MNRAS 424, L44

\item Deason A.J., Belokurov V., Evans N.W., et al., 2012b, MNRAS 425, 2840

\item Deason A.J., Fattahi A., Frenk C.S., et al., 2020, preprint (arXiv: 2002.09497)

\item Dehnen W., Binney J., 1998, MNRAS  294, 429

\item Deng L.-C., Newberg H.J., Liu C., et al., 2012, Res. Astron. Astrophys. 12, 735

\item Drlica-Wagner A., Bechtol K., Mau S., et al., 2020, ApJ 893, Issue 1, id.47

\item Eadie G.M., Harris W.E., and Widrow L.M., 2015, ApJ 806, 54

\item Eadie G.M., Juri\'c M., 2019, ApJ 875, 159

\item Fardal M.A.,  van der Marel R.P.,  Law D.R., et al., 2019, MNRAS 483, 4724

\item Fritz T.K., Battaglia G., Pawlowski M.S., et al., 2018, A\&A 619, 103

\item Fritz T.K., Carrera R., Battaglia G., and S. Taibi, 2019, A\&A 623, 129

\item Fritz T.K., Di Cintio A., Battaglia G., et al., 2020, MNRAS 806, 54  

\item Gibbons S.L.J., Belokurov V., and Evans N.W., 2014, MNRAS 445, 3788

\item Gnedin O.Y., Brown W.R., Geller M.J., and Kenyon  S.J., 2010, ApJ 720, L108

\item Gaia Collaboration, Helmi A., van Leeuwen F., McMillan P.J.,
et al., 2018, A\&A 616, 12

\item Helmi A., 2020, preprint (arXiv: 2002.04340)

\item Hernitschek N., Sesar B., Rix H.-W., et al., 2017, ApJ 850, 96

The HIPPARCOS and Tycho Catalogues,1997, ESA SP--1200

\item Huang Y., Liu X.-W., Yuan H.-B., et al., 2016, MNRAS 463, 2623

\item Irrgang A., Wilcox B., Tucker E., and Schiefelbein L., 2013, A\&A 549, 137

\item Kafle R.R., Sharma S., Lewis G.F., and Bland-Hawthorn J., 2012, ApJ, 761, 98

\item Kallivayalil N., van der Marel R.P., Besla G., et al., 2013, ApJ 764, 161

\item Gaia Collaboration, Kallivayalil N., Sales L.V., Zivick P., et al., 2018, ApJ 867, 19

\item Karachentsev I.D., Kashibadze O.G., Makarov D.I., and R.B. Tully, 2009,
MNRAS 393, 1265

\item Gaia Collaboration, Lindegren L., Hernandez J., Bombrun A., et al., 2018,
A\&A 616, 2  

\item Longeard N., Martin N., Starkenburg E., et al., 2020, MNRAS 491, 356

\item Van der Marel R.P., 2015, Proc. IAU Symp. No. 311, ed. M. Cappellari, and
S. Courteau, 1

\item Massari D., Helmi A., 2018, A\&A 620, 155

\item Massari D., Koppelman H.H., and Helmi A., 2020, A\&A 630L, 4

\item McConnachie A.W., 2012, AJ 144, 4

\item McMillan P.J., 2011, MNRAS 414, 2446

\item Miyamoto M., Nagai R., 1975, PASJ 27, 533

\item  Nadler E.O., Wechsler R.H., Bechtol K., et al., 2020, ApJ 893, Issue 1, id.48

\item Navarro J.F., Frenk C.S., and White S.D.M., 1997, ApJ 490, 493

\item Newton O., Cautun M., Jenkins A., et al., 2018, MNRAS 479, 2853

\item Pace A.B., Li T.S., 2019, ApJ 875, 77

\item Pace A.B., Kaplinghat M., Kirby E., et al., 2020, preprint (arXiv: 2002.09503)

\item Patel E., Besla G., Mandel K., and Sohn S.T., 2018, ApJ  857, 78

\item Pato M., Iocco F., 2015, ApJ 803L, 3

\item Pawlowski M.S., Kroupa P., 2013, MNRAS 435, 1928

\item Pawlowski M.S., Kroupa P., 2013, MNRAS 435, 2116

\item Pojmanski G., 2002, Acta Astron. 52, 397  

Posti L., Helmi A., 2019, A\&A 621, 56

\item Gaia Collaboration, Prusti T., de Bruijne J.H.J., Brown A.G.A., et al., 2016, A\&A 595, A1  

\item Pryor C., Piatek S., and Olszewski E.W., 2010, AJ 139, 839

\item Pryor C., Piatek S., and Olszewski E.W., 2015, AJ 149, 42

Sch\"{o}nrich R., Binney J., and Dehnen W., 2010, MNRAS 403, 1829

\item Simon J., 2018, ApJ 863, 89

\item Sofue Y., 2012, PASJ 64, 75

\item Sofue Y., 2017, PASJ 69, 1

\item Udalski A., Kubiak M., and Szyma$\acute{n}$ski M., 1997, Acta Astron. 47, 319  

\item Wang W., Han J., Cautun M., et al. 2020, SCPMA 63, Issue 10, id.109801

\item Watkins L.L., Evans N.W., and An J.H., 2010, MNRAS 406, 264

\item Xue X.X., Rix H.W., Zhao G., et al., 2008, ApJ 684, 1143

\item Yanny B., Rockosi C., Newberg H.J.,  et al., 2009, AJ 137, 4377

\end{itemize}

\end{document}